\documentclass[3p, sort&compress]{Article_website}
\usepackage{moreverb,url}
\usepackage[colorlinks,bookmarksopen,bookmarksnumbered,citecolor=red,urlcolor=red]{hyperref}
\newcommand\BibTeX{{\rmfamily B\kern-.05em \textsc{i\kern-.025em b}\kern-.08em
T\kern-.1667em\lower.7ex\hbox{E}\kern-.125emX}}

\usepackage{graphicx}			
\usepackage{color}
\usepackage{array}
\usepackage{colortbl}
\usepackage{amsmath}
\usepackage{mathtools}
\usepackage{units}
\usepackage{epstopdf}
\usepackage{stfloats}
\usepackage{titlesec}
\usepackage[normalem]{ulem}

\titleformat*{\section}{\large\bfseries}
\titleformat*{\subsection}{\normalsize\bfseries}

\begin{document}
\title{Combustion regimes in sequential combustors: Flame propagation and autoignition at elevated temperature and pressure}

\author[rvt]{O. Schulz\corref{cor1}}
\ead{oschulz@ethz.ch}
\author[rvt]{N. Noiray\corref{cor1}}
\ead{noirayn@ethz.ch}

\cortext[cor1]{Corresponding author}
\address[rvt]{CAPS Laboratory, Department of Mechanical and Process Engineering,\\ ETH Zurich, 8092 Zurich, Switzerland}

\begin{abstract}
This numerical study investigates the combustion modes in the second stage of a sequential combustor at atmospheric and high pressure. The sequential burner (SB) features a mixing section with fuel injection into a hot vitiated crossflow. Depending on the dominant combustion mode, a recirculation zone assists flame anchoring in the combustion chamber. The flame is located sufficiently downstream of the injector resulting in partially premixed conditions. First, combustion regime maps are obtained from 0-D and 1-D simulations showing the co-existence of three combustion modes: autoignition, flame propagation and flame propagation assisted by autoignition. These regime maps can be used to understand the combustion modes at play in turbulent sequential combustors, as shown with 3-D large eddy simulations (LES) with semi-detailed chemistry. In addition to the simulation of  steady-state combustion at three different operating conditions,  transient simulations are performed: (i) ignition of the combustor with autoignition as the dominant mode, (ii) ignition that is initiated by autoignition and that is followed by a transition to a propagation stabilized flame, and (iii) a transient change of the inlet temperature (decrease by $150\,\text{K}$) resulting into a change of the combustion regime. These results show the importance of the recirculation zone for the ignition and the anchoring of a propagating type flame. On the contrary, the autoignition flame stabilizes due to continuous self-ignition of the mixture and the recirculation zone does not play an important role for the flame anchoring.
\end{abstract}

\maketitle

\section{Introduction} 
Constant-pressure sequential combustion systems \cite{Pennell2017}, or axial staging concepts \cite{Karim2017}, provide significantly higher operational and fuel flexibility compared to conventional single stage combustors. These new type of combustor architectures have emerged in recent years in very large gas turbines (above 500 MW electrical output in single cycle) in order to respond to the needs for fast compensation of inherently-intermittent renewable sources, and for machines which can burn natural gas (NG), but also syngas and hydrogen-enriched NG. In such systems, combustion takes place in two successive stages at constant pressure. The first one features ``classical" lean turbulent premixed flames, usually swirled. The second stage flame results from fuel injection into air-diluted \cite{Pennell2017} or non-diluted \cite{Karim2017} hot gases produced by the lean flame of the first stage. An increasing number of studies dealing with second stage combustion have been published recently. For instance, experimental works investigated the flame stabilization mechanism of non-premixed \cite{Sullivan2014,Sidey2015,Fleck2013,Panda2016} and premixed \cite{Kolb2015a,Wagner2017} reactive jet in hot crossflow configurations. In their respective configurations, \citet{Sullivan2014} and \citet{Fleck2013} proposed autoignition as the dominant flame stabilization mechanism, while \citet{Micka2012b}, and \citet{Wagner2017} identified a mix of autoignition and flame propagation stabilizing the flame. Premixed and non-premixed reactive jets in vitiated crossflows were also investigated with large eddy simulation (LES), e.g. \cite{SchulzASME,SchulzRJICF,Weinzierl2016} and direct numerical simulations (DNS), e.g. \cite{Kolla2012,Grout2012,Minamoto2015}, giving further insight in the associated complex stabilization mechanisms. These examples feature a flame that is stabilized close to the fuel injection by partially-premixed combustion \cite{Kolla2012,Grout2012}, differential diffusion initiating chemical reactions \cite{Minamoto2015} or mix between premixed and partially-premixed combustion \cite{SchulzASME,SchulzRJICF}, and they display a similar configuration as the axial staging concept presented in \cite{Karim2017}.

Another type of sequential combustor \cite{Pennell2017} features a flame that is stabilized substantially downstream of the fuel injector in order to ensure good mixing of the fuel with the vitiated flow before the former is consumed, and it falls into the category of partially-premixed combustion. The response of these flames to different types of perturbation, such as acoustic velocity \cite{Zellhuber2014,yang2015} or temperature fluctuations \cite{Schulz2018,Scarparto_2016,Bothien2017} have been investigated numerically in the context of thermoacoustic instabilities in sequential combustors \cite{SchulzSYMP,Berger2018}.

The operation of current practical sequential burners (SB), and the development of future SBs would benefit from an improved understanding of the combustion regimes that can exist in these systems. Recently, the existence of both autoignition-based and propagation-based flame anchoring in a generic SB operated at atmospheric pressure was demonstrated experimentally \cite{Ebi2018}. This study follows up on a numerical investigation of the same combustor highlighting the importance of the combustion mode on the flame dynamics at atmospheric condition \cite{Schulz2018}. In the latter reference, as well as in \cite{Habisreuther2013,Krisman2018a}, the transition between autoignition and propagation has been investigated for one dimensional (1-D) laminar flames. \citet{Habisreuther2013} showed that when the temperature of methane-air mixtures  is increased well above the autoignition temperature, the flame structure significantly changes and the reactants consumption speed increases. They also formulate a limiting criterion for the onset of this effect, which is based on autoignition delay time, laminar flame speed and residence time of the mixture before the ``flame foot'' position. These findings were confirmed by \citet{Schulz2018} for a range of mixture fractions and reactants residence times, and the increase of burning velocity was linked to the heat release rate associated with autoignition reactions upstream of the flame front. These conclusions have also been drawn by \citet{Krisman2018a} for ethanol- and dimethyl ether-air flames. The above mentioned numerical investigations consider steady-state 1-D flames \cite{Habisreuther2013,Schulz2018,Krisman2018a} or steady-state operation at atmospheric condition of a generic SB \cite{Schulz2018}. In the present numerical work, the combustion regimes of both steady-state \emph{and} transient operation (ignition and change of operating conditions) are investigated in order to complement the experimental work from \citet{Ebi2018}. Also, the present LES investigation (with semi-detailed chemistry) aims at bringing a deeper understanding of the combustion regimes at atmospheric \emph{and} high pressure ($10\,\text{bar}$) in this generic SB configuration.

In general, autoignition and flame propagation are two different mechanisms that can stabilize flames in high-temperature flows. Lifted flames have been largely investigated for various configurations experimentally \cite{Cabra2005,Gordon2008,Arndt2016a,Wagner2017b,Macfarlane2018} and numerically \cite{Cabra2005,Gordon2007a,Yoo2009,Kerkemeier2013,Karami2015,Deng2015,Minamoto2016a,Schulz2017,SchulzRJICF}. The main anchoring mechanism is sometimes attributed to autoignition (e.g. \cite{Cabra2005,Gordon2007a,Gordon2008,Yoo2009,Kerkemeier2013,Arndt2016a,Schulz2017,Macfarlane2018}), in other situations to flame propagation (e.g. \cite{Deng2015,Karami2015}), to a mix between the two depending on the flame branch (e.g. \cite{Wagner2017b,SchulzRJICF}), or to flame propagation enhanced by autoignition \cite{Minamoto2016a}. Considering the mixing of two or more streams in homogeneous 0-D reactors, the shortest autoignition time is obtained for the most reactive mixture fraction $Z_{\,\text{mr}}$ in the mixture fraction space $Z$ (see for instance  \cite{Mastorakos1997}). In many cases (e.g. \cite{Cabra2005,Schulz2017}), $Z_{\,\text{mr}}$ corresponds to very lean compositions, and the associated flow regions, where induction reactions and autoignition take place, are difficult to identify experimentally because fuel concentration and heat release rate are very low \cite{Arndt2016a,SchulzRJICF,Wagner2017b}.

For turbulent flows, \citet{Mastorakos1997} highlighted another important property for autoignition: the scalar dissipation rate $\chi$. High scalar dissipation rates correspond to large gradients of $Z$ and therefore large heat losses. Autoignition occurs preferentially at low $\chi$ conditioned on the most reactive mixture fraction as shown in many subsequent studies, for example \cite{Cao2007,Krisman2017, SchulzRJICF, Schulz2018}. Regions with low $\chi$ can occur for example in the core of vortices, as shown, for example, with DNS simulations in \cite{Sreedhara2002}. Other studies, however, also observed an acceleration of autoignition due to turbulence and molecular transport. For a review about non-premixed autoignition, one can refer to \cite{Mastorakos2009}.

The goal of this work is to provide a deeper understanding of the combustion regimes in sequential combustors. As a first step, 0-D and 1-D simulations are performed in order to grasp the key conditions leading to autoignition or front propagation for a given autoignitive mixture, under homogeneous and turbulence-free conditions. In section \ref{sec:2} the effect of vitiated hot gas temperature is further investigated and the capability of the chemical explosive mode analysis (CEMA) \cite{Lu2010} to distinguish between autoignition and flame propagation at elevated temperatures is scrutinized. Finally, section \ref{sec:2} presents a transient 1-D simulation of the combustion of an autoignitive methane-air mixture with a transition from autoignition to propagation. In section \ref{sec:sequ_comb}, 3-D LESs of the sequential combustor are performed. The section starts with the introduction of the geometry and the numerical methods. Then, the steady-state combustion process at three operating conditions is presented and discussed in the light of the results from the 1-D simulations of section \ref{sec:2}. Finally, the transient ignition sequence for two of these operating conditions, which are respectively characterized by different dominant steady-state combustion regimes (autoignition and propagation). A simulation with a transient change of operating condition resulting in a change of the combustion regime is also presented.

\section{Autoignition and propagation at elevated gas temperatures} \label{sec:2}
\subsection{Combustion regime maps obtained from 0-D reactor and 1-D flame simulations}
\begin{figure*}[!t]
	\begin{center}
	\includegraphics[width=150mm]{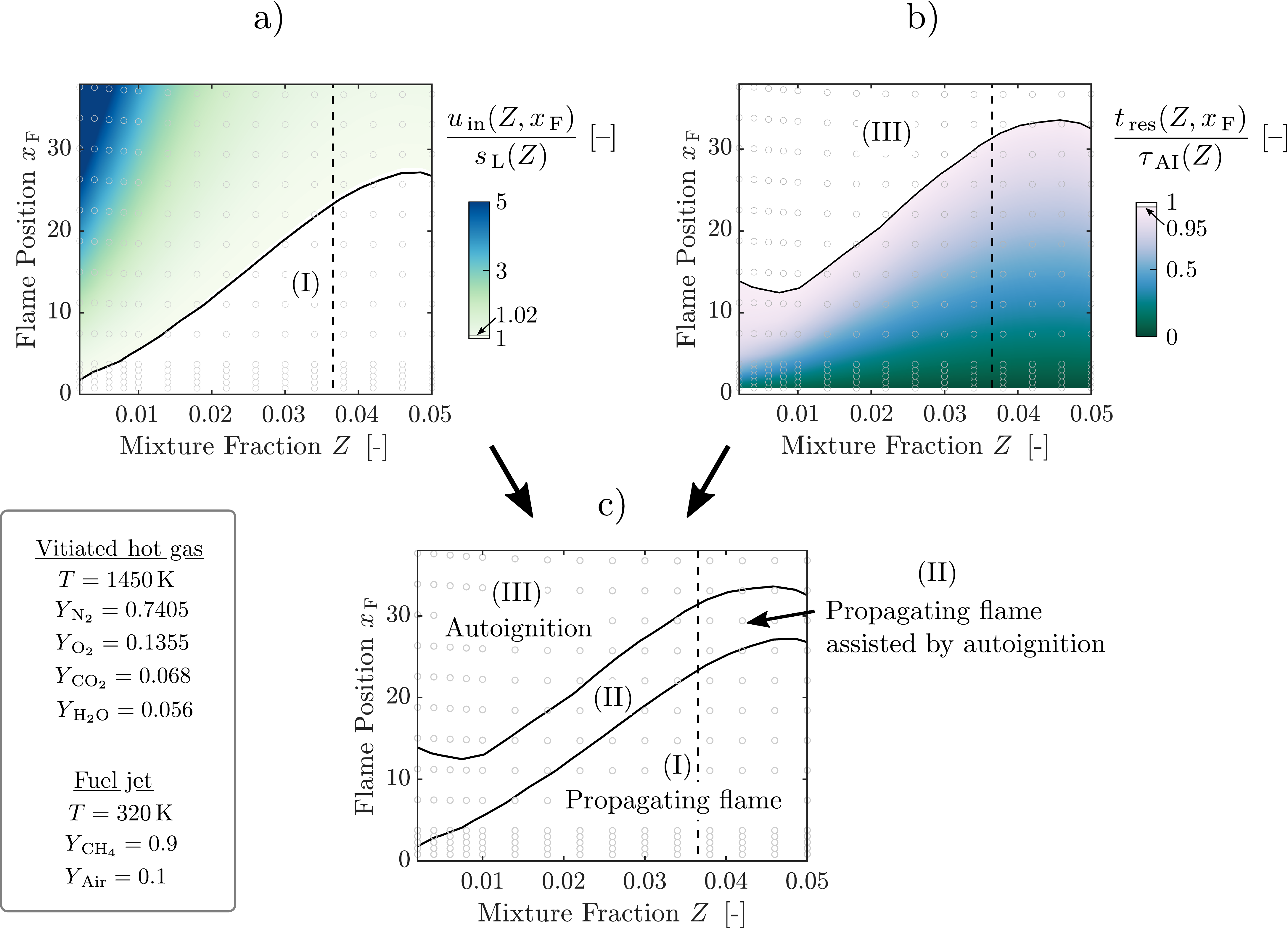}
	\caption{a) Normalized inlet velocity $u_\text{in}(Z,x_{\,\text{F}})$ for converged 1-D steady flame solutions (obtained from Cantera) with mixture fraction $Z$ and a flame front located at $x_\text{F}$. The inlet velocity $u_\text{in}(Z,x_{\,\text{F}})$ is normalized by the velocity at the shortest flame position $u_{\,\text{in}}(Z,x_{\,\text{F,min}})$, which is assumed to be the laminar flame speed $s_\text{L}(Z)$. b) Reactants residence time $t_\text{res}(Z,x_\text{F})\simeq x_\text{F}/u_\text{in}(Z,x_\text{F})$ of the same 1-D flame solutions as in a), normalized by the autoignition delays $\tau_\text{AI}(Z)$, which were obtained from 0-D reactor simulations. c) Combustion regime diagram deduced from a) and b). Pressure: $1\,\text{bar}$.}
	\label{fig:map1450K}	
	\end{center}
\end{figure*}
In this section, 1-D flame simulations using Cantera \cite{cantera} with the detailed GRI-Mech 3.0 mechanism \cite{gri30} are performed in order to investigate the conditions for which autoignition and flame propagation respectively govern the combustion process at atmospheric pressure. Several simulations are performed for a range of mixture fraction $Z$, which describes the mixing between two streams: (i) air-diluted products from a lean ($\phi=0.75$) methane-air flame at different temperatures (in this section: $1000\,$, 1450 and  $1600\,\text{K}$), and (ii) ``cold'' methane at $320\,\text{K}$. More details about the gas composition are given in Fig. \ref{fig:map1450K}.

Figure \ref{fig:map1450K} shows two contour plots (top) that were used to obtain a combustion regimes map (bottom) for varying flame position $x_{\,\text{F}}$ and mixture fraction $Z$. Each circle represents one 1-D steady flame solution computed with Cantera. To compute each of these solutions, the required input parameters are the pressure, the mixing temperature and the species composition at the inlet of the domain. For a given domain length $L_{\,\text{d}}$ and a given inlet mixture fraction $Z$, Cantera provides a unique steady flame solution where the flame is located between one fourth and half of domain length $L_{\,\text{d}}$. This eigenvalue-problem solution is the spatial distribution of the mixture velocity, temperature and composition obtained from the detailed chemical scheme. Upstream of the flame front, the mixture velocity is close to the inlet velocity $u_\text{in}$. For each mixture fraction considered, the length $L_{\,\text{d}}$ was varied between $2$ and $100\,\text{mm}$ in order to get solutions for several flame positions $x_{\,\text{F}}$.

For a freely propagating flame $u_{\,\text{in}}$ equals the laminar flame speed $s_{\,\text{L}}$. In Fig. \ref{fig:map1450K}a the inlet velocity $u_{\,\text{in}}(Z,x_\text{F})$ is normalized by the laminar flame speed $s_{\,\text{L}}(Z)$. Here,  $s_{\,\text{L}}$ is assumed to be the inlet velocity for short domain solutions, where the residence time of the reactants is so short that autoignition chemistry does not significantly influence the reactants consumption speed, i.e. $s_L(Z)\approx u_{\,\text{in}}(Z,x_\text{F,min})$. One can define a ``flame propagation'' region (I) in which the increase of the reactants consumption speed $u_\text{in}(Z,x_\text{F})$ from the laminar flame speed $s_\text{L}(Z)$ does not exceed 2$\%$ of $s_\text{L}(Z)$; this region is colored in white. Out of region (I) the influence of autoignition chemistry can significantly contribute to the reactants consumption speed. As a consequence, a unique flame speed does not exist anymore and the flame stabilization velocity depends on the configuration, for example, as considered in this work, the position where the flame stabilizes. Flames away from region (I) stabilize at inlet velocities that are substantially larger than $s_{\,\text{L}}$ (up to 5 times larger), which is in agreement with \cite{Schulz2018,Krisman2018a}.

\begin{figure}
	\begin{center}
	\includegraphics[width=88mm]{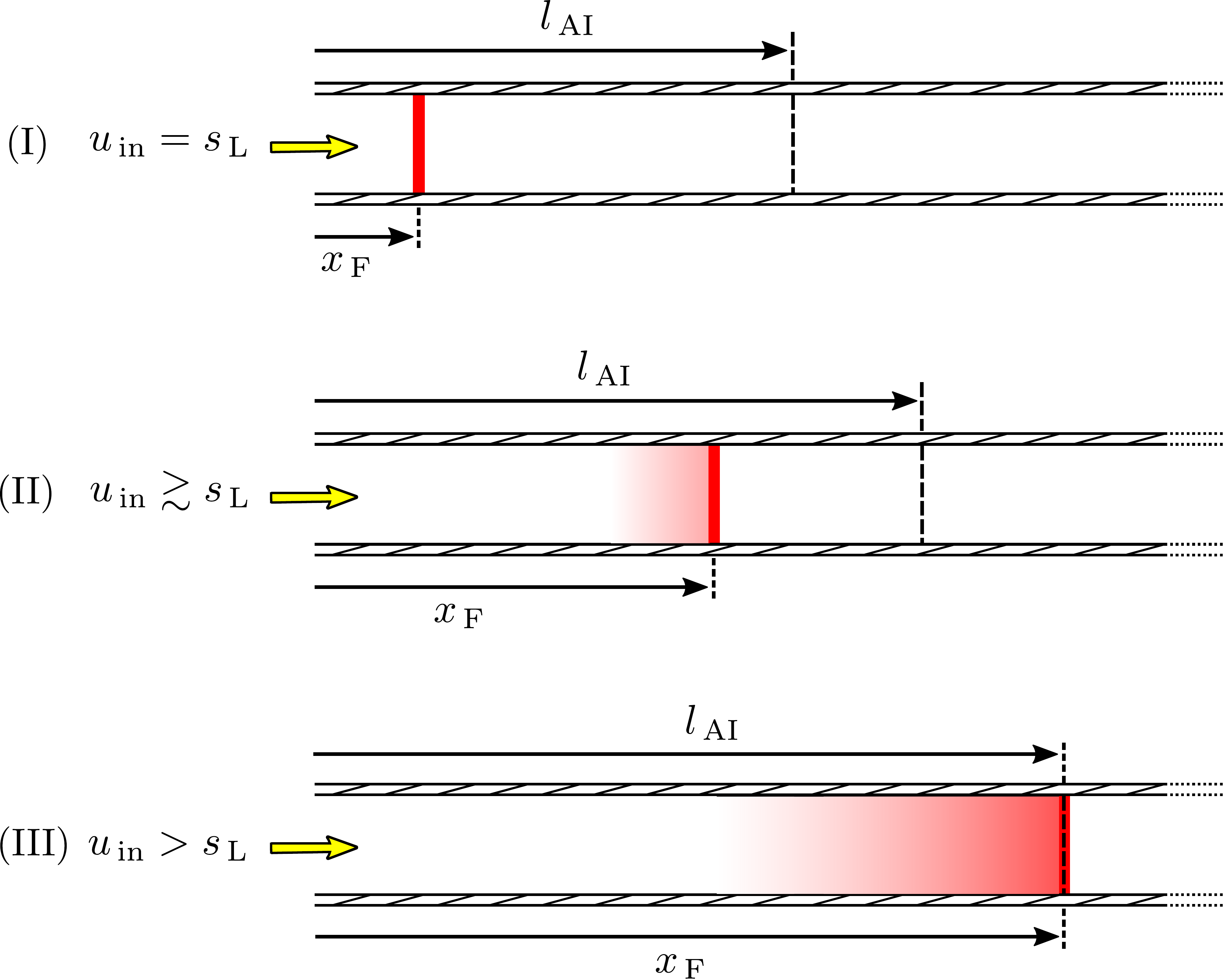}
	\caption{Sketch of three 1-D steady-state flames located in (I) -- pure propagating flame, (II) -- propagating flame assisted by autoignition and (III) -- autoignition flame (compare with Fig. \ref{fig:map1450K}). The red contour represents heat release rate $\dot{q}$.}
	\label{fig:sketch_flames}
	\end{center}
\end{figure}
\begin{figure*}[ht!]
	\begin{center}
	\includegraphics[width=160mm]{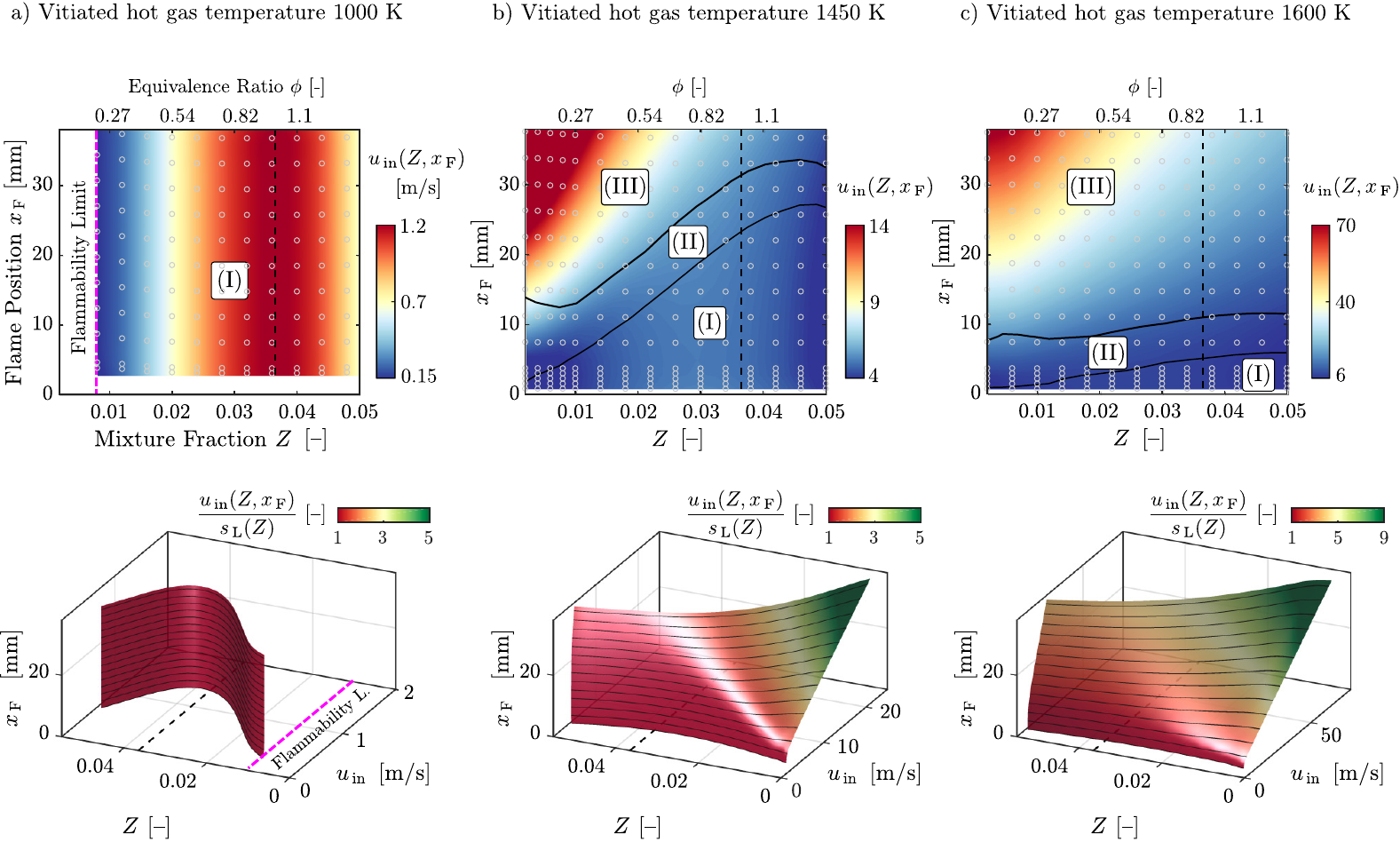}
	\caption{The effect of the vitiated hot gas temperature on combustion regimes maps obtained from 1-D Cantera simulations. Top: contours of inlet velocity of 1-D flame simulations $u_{\,\text{in}}(Z,x_{\,\text{F}})$ for three vitiated hot gas temperatures: (a) $1000\,\text{K}$, (b) $1450\,\text{K}$ and (c) $1600\,\text{K}$. The lines highlight the regimes boundaries that are shown in Fig. \ref{fig:map1450K}. The vertical black dashed lines indicate the stoichiometric condition. Bottom: The same data are presented as a 3-D surface, which gives the ``coordinates'' of steady flames solutions. This surface is colored by $u_{\,\text{in}}(Z,x_{\,\text{F}})/s_{\,\text{L}}(Z)$. Pressure: $1\,\text{bar}$.}
	\label{fig:regimes1}	
	\end{center}
\end{figure*}
Figure \ref{fig:map1450K}b shows the residence time of the reactants normalized by the autoignition delay $\tau_{\,\text{AI}}(Z)$. The reactants residence time was computed by integrating the velocity from the inlet to the flame position $x_{\,\text{F}}$: $t_{\,\text{res}}(Z,x_\text{F})=\int_0^{x_{\,\text{F}}}dx/u\,\simeq x_{\,\text{F}}/(u_{\,\text{in}}(Z,x_{\,\text{F}}))$. The autoignition delays $\tau_{\,\text{AI}}(Z)$ were computed with 0-D Cantera reactor simulations. The autoignition time is defined at the highest temperature gradient throughout the paper. An ``autoignition'' region (III) is defined in which the reactants residence time is $\geq95\%$ of $\tau_{\,\text{AI}}(Z)$ (white region in contour plot). A combined representation of the two zones is shown in Fig. \ref{fig:map1450K}c. In between (I) and (III), another zone (II) is defined. It represents propagating flames that are assisted by autoignition chemistry upstream of the high heat release rate flame front. In \cite{Schulz2018}, a comparison of two 1-D flames which were located in (I) and (II) was presented. It was shown that the heat release rate $\dot{q}$ sharply increases across the flame front for the regime (I). For the flame in (II), a moderate increase of $\dot{q}$ was already observed upstream of the flame front leading to an increase of the reactants temperature in this region.

Figure \ref{fig:sketch_flames} presents sketches of three flames located in (I), (II) and (III). The same $Z$ and, therefore, also the same equivalence ration $\phi$ was assumed for the three flames. (I) shows a pure propagating flame without any heat release rate $\dot{q}$ upstream of the flame front. For (II), a stable flame is established with an inlet velocity that is slightly higher than $s_{\,\text{L}}$. In this case, the theoretical autoignition length $l_{\,\text{AI}}=u_{\,\text{in}}\tau_{\,\text{AI}}$ is therefore located downstream compared to (I). The red hue indicates a moderate $\dot{q}$ upstream of $x_{\,\text{F}}$. For (III), the front position $x_{\,\text{F}}$ is defined by the autoignition length $l_{\,\text{AI}}$, i.e. the reactants residence time $x_{\,\text{F}}/u_{\,\text{in}}$ equals the autoignition time $\tau_{\,\text{AI}}$.

\begin{figure}
	\begin{center}
	\includegraphics[width=88mm]{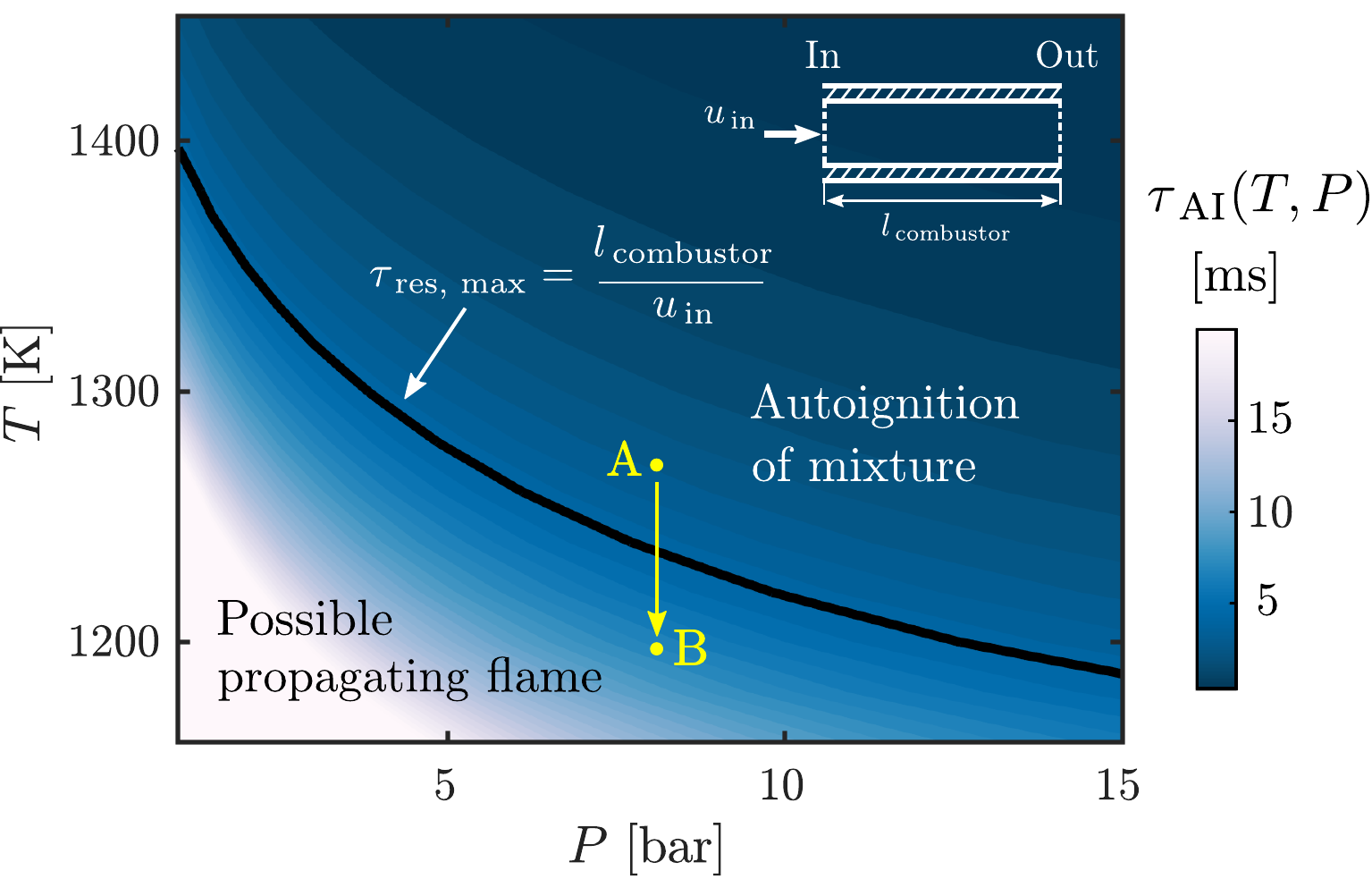}
	\caption{Contour of autoignition time $\tau_{\,\text{AI}}(T,P)$ for varying temperature $T$ and pressure $P$ obtained from 0-D reactor simulations. Black line at the maximum residence time $\tau_{\,\text{res, max}}$ of a burner shown in a simplified sketch in the top right corner. For $\tau_{\,\text{AI}}(T,P)<\tau_{\,\text{res, max}}$ the mixture auto-ignites; for $\tau_{\,\text{AI}}(T,P)>\tau_{\,\text{res, max}}$ the mixture does not auto-ignite but a propagating flame can exists (``B") after autoignition of the mixture (``A").}
	\label{fig:map_tau}
	\end{center}
\end{figure}
Figure \ref{fig:regimes1} presents inlet velocity $u_{\,\text{in}}(Z,x_{\,\text{F}})$ maps with regime boundaries (top) and 3-D representations of $u_{\,\text{in}}(Z,x_{\,\text{F}})/s_{\,\text{L}}(Z)$ (bottom) for three air-diluted lean flame product temperatures: (a) $1000\,\text{K}$, (b) $1450\,\text{K}$ and (c) $1600\,\text{K}$. As in Fig. \ref{fig:map1450K}, the axis in the top row are flame position $x_{\,\text{F}}$ and mixture fraction $Z$. Each circle represents a steady 1-D Cantera flame solution. The contour colorbar scale is changed from (a) to (c). For (a), only freely propagating flame solutions were obtained (regime (I)). As shown in the top and bottom, for any $x_{\,\text{F}}$ at constant $Z$, the inlet velocity does not change, and therefore, a unique laminar flame speed $s_{\,\text{L}}(Z)=u_{\,\text{in}}(Z)$ is determined. 
\begin{figure*}[!t]
	\begin{center}
	\includegraphics[width=120mm]{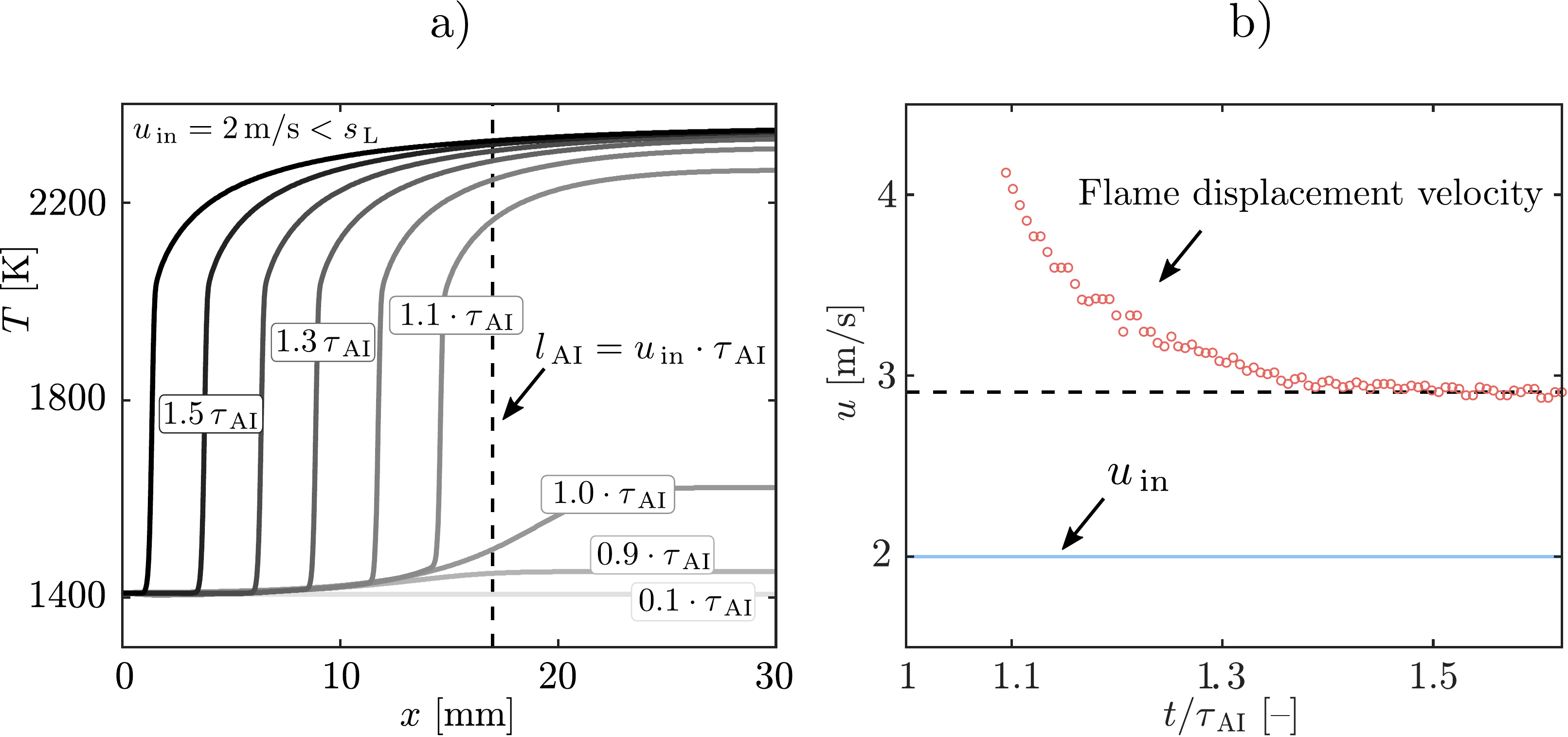}
	\caption{a) Temperature profiles of a transient 1-D simulation at varying instants of time. The mixture auto-ignites at $\tau_{\,\text{AI}}=8.5\,\text{ms}$, and, subsequently, propagates towards the domain inlet. b) Transient flame displacement velocity $u_{\,\text{d}}$ obtained from the same simulation as in (a). Species composition ($\phi=1$): $Y_{\,\text{CH}_4}=0.033$, $Y_{\,\text{N}_2}=0.716$, $Y_{\,\text{O}_2}=0.131$, $Y_{\,\text{H}_2\text{O}}=0.054$ and $Y_{\,\text{CO}_2}=0.066$.}
	\label{fig:T_over_T_t}
	\end{center}
\end{figure*}
For $Z<0.008$, Cantera did not converge to a solution reaching the flammability limit. As the vitiated hot gas temperature is increased to $1450\,\text{K}$ (same case as in Fig. \ref{fig:map1450K}), autoignition chemistry can contribute to the reactants consumption and the three regimes can coexist depending on $x_{\,\text{F}}$ and $Z$. The onset of autoignition is visible where the 3-D iso-surface, which gives the ``coordinates'' of steady flame solutions, bends towards higher inlet velocities. This effect manifests itself in a more pronounced manner for increasing $x_{\,\text{F}}$ at small $Z$. As the vitiated hot gas temperature is further increased to $1600\,\text{K}$, only a small region (I) with pure flame propagation is identified. Reactants residence times approach $\tau_{\,\text{AI}}$ even for flame positions that are very close to the domain inlet, and fast autoignition of the mixtures dominates allowing flame stabilization at very high inlet velocities (up to $70\,\text{m/s}$).

\subsection{The transient evolution from autoignition to propagation}
Figure \ref{fig:map_tau} shows the contour of autoignition times $\tau_{\,\text{AI}}(T,P)$ for varying temperature $T$ and pressure $P$. These results were derived from 0-D reactor simulations using Cantera with a species composition that is characteristic for the investigated burner of this study ($\phi=0.4$). For each condition, the mixture has the property to eventually auto-ignite after $\tau_{\,\text{AI}}$. However, in the simplified configuration sketched in the upper part of Fig. \ref{fig:map_tau}, a maximum residence time $\tau_{\,\text{res, max}}$ that depends on the combustor length $l_{\,\text{combustor}}$ and the inlet velocity $u_{\,\text{in}}$ can be defined, and it is here exemplified with the black line. For $\tau_{\,\text{AI}}(T,P)<\tau_{\,\text{res, max}}$ (dark contour), the unburnt mixture auto-ignites in the combustor. Depending on $u_{\,\text{in}}$, the resulting flame can be stabilized by autoignition or evolve into a propagating flame, as shown with a transient 1-D flame simulation in the next paragraph. For $\tau_{\,\text{AI}}(T,P)>\tau_{\,\text{res, max}}$ (light contour), the mixture does not ignite in the domain, similar to the ``no ignition" regime reported by \citet{Markides2005}. However, one could imagine the following scenario: the mixture auto-ignites (marked with an ``A" in Fig. \ref{fig:map_tau}) and, subsequently, boundary conditions are changed such that $\tau_{\,\text{AI}}(T,P)>\tau_{\,\text{res, max}}(T,P)$, for example by decreasing the inlet temperature (marked with a ``B"). This would result in a propagating type flame that was initiated by autoignition. Such a transient change of boundary conditions resulting in a shift of the dominant combustion regime is presented in subsection \ref{sec:transient} for the 3-D sequential combustor.

 \begin{figure*}[t!]
	\begin{center}
	\includegraphics[width=160mm]{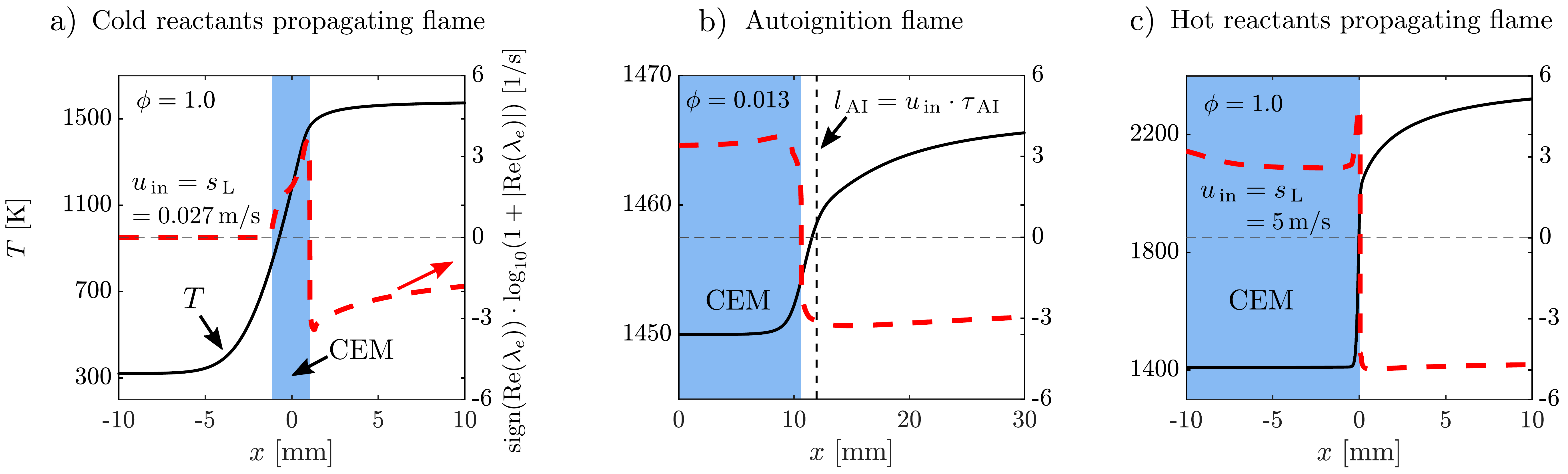}
	\caption{Profiles of temperature and order of magnitude of the growth rate of the least stable chemical mode for three 1-D flames. Positive real part of eigenvalues Re($\lambda_e$) indicate chemical explosive modes (CEM) highlighted in blue. (a) propagating flame without CEM in unburnt reactants. b) autoignition flame with CEM in unburnt reactants. (c) propagating flame with CEM in unburnt reactants. Species composition in (a) and (c): $Y_{\,\text{CH}_4}=0.033$, $Y_{\,\text{N}_2}=0.716$, $Y_{\,\text{O}_2}=0.131$, $Y_{\,\text{H}_2\text{O}}=0.054$ and $Y_{\,\text{CO}_2}=0.066$. Species composition in (b): $Y_{\,\text{CH}_4}=0.0005$, $Y_{\,\text{N}_2}=0.74$, $Y_{\,\text{O}_2}=0.1354$, $Y_{\,\text{H}_2\text{O}}=0.056$ and $Y_{\,\text{CO}_2}=0.068$. $P =1\,\text{bar}$.}
	\label{fig:CEMA}	
	\end{center}
\end{figure*} 
Figure \ref{fig:T_over_T_t} shows the transient evolution from an auto-igniting unburnt mixture to a propagating flame. The unburnt gas has a temperature of 1400 K at stoichiometric equivalence ratio. The 1-D simulation was performed with the time advancement solver AVBP. The velocity at the inlet of the computational domain (x = 0 mm) was fixed at $u_{\,\text{in}}=2\,\text{m/s}$. The starting point of the simulation was at $t=0$. Figure \ref{fig:T_over_T_t}a shows temperature profiles at different instants of time. The unburnt mixture has the property to auto-ignite at the autoignition time $\tau_{\,\text{AI}}=8.5\,\text{ms}$, which was computed with a 0-D reactor simulation. The autoignition length $l_{\,\text{AI}}$ is marked with a dashed vertical line. Indeed, one can observe a rapid increase of temperature $T$, as shown at instants from $0.9\,\tau_{\,\text{AI}}$ to $1.1\,\tau_{\,\text{AI}}$. Afterward, from $1.1\,\tau_{\,\text{AI}}$ to $1.5\,\tau_{\,\text{AI}}$, the flame propagates against the incoming flow. The inlet velocity $u_{\,\text{in}}=2\,\text{m/s}$ is smaller than the laminar flame speed $s_{\,\text{L}}=5\,\text{m/s}$, which was computed by a 1-D Cantera simulation with very short domain length. This leads to a continuous flame displacement towards the inlet of the domain. The flame displacement velocity $u_{\,\text{d}}$ defined as the flame front speed relative to the flow is shown in Fig. \ref{fig:T_over_T_t}b starting from $1.1\,\tau_{\,\text{AI}}$. Close to $l_{\,\text{AI}}$, the displacement speed reaches values of $4\,\text{m/s}$. Here, the sum of displacement speed and inlet velocity is higher than $s_{\,\text{L}}$ due to autoignition chemistry in the unburnt gases, corresponding to the ``flame propagation assisted by autoignition" regime (II) in Fig. \ref{fig:map1450K}. Consequently, also the fresh gas temperature upstream of the pre-ignition zone increases, which is shown in Fig. \ref{fig:T_over_T_t}a. From the instant $t\approx1.4\,\tau_{\,\text{AI}}$, the displacement speed approaches a constant value of $\approx 3\,\text{m/s}$ and, therefore, the sum of $u_{\,\text{d}}$ and $u_{\,\text{in}}$ equals $s_{\,\text{L}}$. Here, the residence time of the unburnt gases is short enough such that the effect of autoignition chemistry vanishes upstream of the flame front corresponding to regime (I). These results show that all three regimes can exist in reactants at elevated temperatures.

\subsection{Chemical explosive mode analysis (CEMA)}
CEMA was developed by Lu et al. \cite{Lu2010} as a diagnostic tool to identify flame and ignition structures, and has been applied in many studies, such as \cite{Shan2012,Luo2012,Deng2015,SchulzRJICF}. In the present study, we investigate the capability of CEMA to detect autoignition and propagation type flames for unburnt mixtures at elevated temperatures.

Local species concentrations and temperature were given as an input to pyJac \cite{pyJac} that outputs an analytical expression of $J_{\,\omega}$, the Jacobian matrix of the chemical source terms. Chemical explosive modes (CEMs) are defined as the eigenmodes of $J_{\,\omega}$ associated with positive real part eigenvalues $\lambda_{\,\text{e}}$. This means that an infinitesimal small perturbation of this chemical mode would exponentially grow in an adiabatic, isolated environment. A continuation approach was used to track the CEM across the front, where $\text{Re}(\lambda_{\,\text{e}})$ changes sign and the modes associated with energy and element conservations were discarded.

\begin{figure*}[t!]
	\begin{center}
	\includegraphics[width=160mm]{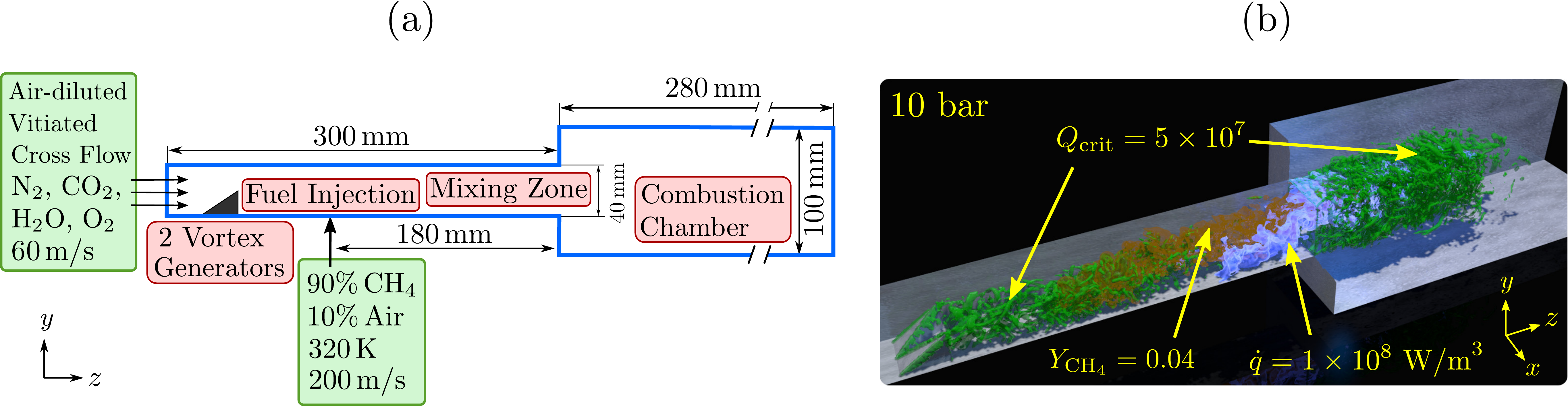}
	\caption{(a): Sketch of the simulated second stage sequential combustion system. (b): Instantaneous iso-contours of $Q$-criterion \cite{Hunt1988} ($5\times 10^7$), methane mass fraction ($Y_{\,\text{CH}_4}=0.04$) and heat release rate ($1\times 10^8\,\text{W/m}^3$) at 10 bar.}
	\label{fig:sketch}
	\end{center}
\end{figure*}
Figure \ref{fig:CEMA} presents profiles of temperature $T$ (left $y$-axis) and the order of magnitude of the growth rate $\text{Re}(\lambda_{\,\text{e}})$ of the least stable chemical mode (right $y$-axis) for three 1-D flames (a) to (c). For the right $y$-axis, the same normalization as in \cite{Luo2012} was used. The propagating flame with cold reactants (a) is characterized by zero $\lambda_{\,\text{e}}$ in the unburnt mixture. The eigenvalue becomes positive in the preheat zone of the flame, which is highlighted with the blue patch. This zone is driven by back diffusion of energy and radicals and eventually ignites. The crossover point of the CEM separates pre- and post-ignition zones. On the contrary, flame (b) is governed by autoignition after a length $l_{\,\text{AI}}$, highlighted with a dashed vertical line. This length depends on the inlet velocity $u_{\,\text{in}}=6.7\,\text{m/s}$ and the autoignition delay $\tau_{\,\text{AI}}=1.79\,\text{ms}$, which was computed with a 0-D reactor simulation. Indeed, an explosive mode is detected in the unburnt reactants, highlighted in blue. For boundary conditions (a) and (b), CEMA can be applied to distinguish between autoignition and propagation type flame fronts. However, at conditions with elevated unburnt gas temperatures, we observed propagating flames with a CEM in the unburnt mixture (Fig. \ref{fig:CEMA}c). The mixture has, therefore, the tendency to auto-ignite (see transient simulation in Fig. \ref{fig:T_over_T_t}). However, after the autoignition event, the flame can propagate against the incoming flow, as captured in the steady state solution in Fig. \ref{fig:CEMA}c. Both regimes can exist in reactants characterized by a CEM, and CEMA would detect an autoignition flame although the flame \emph{propagates} into the CEM reactants. After the submission of this paper, \citet{Xu2018} introduced an updated version of CEMA that besides the chemical source terms also takes into account the effect of the diffusion source terms. This improvement allows CEMA to distinguish between the combustion regimes in the canonical problems presented in Figs. \ref{fig:T_over_T_t} and \ref{fig:CEMA}. The implementation and application of this improved analysis will be the topic of future work.

\section{Application to a generic sequential combustor} \label{sec:sequ_comb} 
\subsection{Configuration and numerical methodology} \label{config}
A sketch of the configuration is shown in Fig. \ref{fig:sketch}a. Only the second stage of the ETH sequential combustor \cite{SchulzSYMP} was simulated. A mixture of combustion products from a perfectly-premixed first stage and of fresh dilution air was imposed at the domain inlet with a bulk velocity of $60\,\text{m/s}$. The composition of this vitiated inlet, which acts as a crossflow for the fuel jet, was the same as in \cite{Schulz2018}: $Y_{\,\text{N}_2}=0.7405$, $Y_{\,\text{O}_2}=0.1355$, $Y_{\,\text{H}_2\text{O}}=0.056$ and $Y_{\,\text{CO}_2}=0.068$. Figure \ref{fig:sketch}b shows a 3-D rendering of the iso-contours of the $Q$-criterion \cite{Hunt1988}, CH$_4$ mass fraction, and heat release rate for an instantaneous snapshot at $10\,\text{bar}$. Three different cases were simulated: (a) vitiated crossflow temperature $T_{\,\text{CF}}$ of $1350\,\text{K}$, and operating pressure $P$ of $10\,\text{bar}$, (b) $T_{\,\text{CF}}$ of $1200\,\text{K}$ and $P$ of $10\,\text{bar}$, and (c) $T_{\,\text{CF}}$ of $1450\,\text{K}$ and $P$ of $1\,\text{bar}$. Fuel (90\% $Y_{\,\text{CH}_4}$ and 10\% $Y_{\,\text{Air}}$) at $320\,\text{K}$ was injected as a jet in crossflow ($d_{\,\text{jet}}=2.6\,\text{mm}$) $180\,\text{mm}$ upstream of the combustion chamber inlet. Two vortex generators ensure an improved mixing between the fuel stream and the hot vitiated crossflow. More details about the simulated cases such as the boundary conditions, the main physical and numerical parameters, and the performed LESs can be found in Table \ref{tab1}; a more detailed description of the configuration was presented in \cite{Schulz2018}. For the present study, minor species, including $\text{NO}_\text{x}$ from the lean combustion process in the first stage, were neglected.
\begin{table*}
    \centering
    \footnotesize
        \caption{Boundary conditions and main physical and numerical parameters for the three simulated operating conditions. The last three rows show the investigated LESs. The integral length scale $l_{\,\text{t}}$ corresponds to $1/5$ of the mixing section height. $<\cdot>$ denotes spatial averaging. $u'_{\,\text{rms}}$ denotes the spatially filtered turbulent intensity which was resolved on the LES mesh. Therefore, $Re_{\,\text{t}}$ is also a spatially filtered quantity.}
    \begin{tabular}{ p{9.0cm} p{1.6cm} p{1.6cm} p{1.6cm} }
    \hline
   Simulation & 10B\_1350K & 10B\_1200K & 1B\_1450K \\
    \hline
    Pressure [bar] & 10 	& 10 & 1 \\ 
    Vitiated crossflow temperature $T_{\,\text{CF}}$ [K] & 1350 & 1200 & 1450 \\ 
    Vitiated crossflow mass flow [g/s] & 240 & 270 & 23 \\ 
    Crossflow velocity $\overline{u}_{\,\text{CF}}$ [m/s] & 60 & 60 & 60  \\
    Fuel mass flow [g/s] & 6.6 	& 6.6 & 0.66 \\ 
    Jet-to-crossflow momentum ratio $(\rho_{\,\text{jet}}\, u_{\,\text{jet}}^2)/(\rho_{\,\text{CF}}\, u_{\,\text{CF}}^2)$ & 27 & 24 & 29 \\
    Thermal power [kW] & 300 & 300 & 30 \\ 
    $Y_{\text{O}_2}$ (O$_2$ from both lean combustion in 1$^\text{st}$ stage and dilution air) [--] & 0.1355 & 0.1355 & 0.1355 \\ 
     Global equivalence ratio $\phi_{\,\text{g}}$ [g/s] & 0.7 & 0.62 & 0.76 \\ 
    Thermal power [kW] & 300 & 300 & 30 \\ 
    Laminar flame speed $S_{\,\text{L}}$ at $\phi_{\,\text{g}}$ and $T_{\,\text{CF}}$ [m/s] & 1.1 & 0.62 & 4.5 \\
    Laminar flame thickness $\delta_{\,\text{L}}$ (propagating flame) at $\phi_{\,\text{g}}$ and $T_{\,\text{CF}}$  [mm] & 0.08 & 0.1 & 0.37 \\
     Turbulence strength spatially-averaged in flame region ${<}u'_{\,\text{rms}}{>}$ [m/s] & 7.8 & 7.8 & 8.0 \\
        Integral length scale $l_{\,\text{t}}$ [mm] & 8 & 8 & 8 \\
    Turbulent Reynolds number $Re_{\,\text{t}}=\dfrac{{<}u'_{\,\text{rms}}{>}\,l_t}{S_{\,\text{L}} \delta_{\,\text{L}}}$ & $709$ & $1006$ & $38$ \\
    Dominant combustion regime & Autoignition & Propagation & Propagation \\
    \hline
    Number of tetrahedral grid cells & $\approx$ 60 million & $\approx$ 74 million & $\approx$ 16 million\\
    Thermal resistance $R$ in mixing section [K/W] & 0.005 & 0.005 & 0.02 \\
    $R$ in combustion chamber [K/W] & 0.01 & 0.01 & 0.04\\
    \hline
    LES of the ignition sequence & $\times$ & - & $\times$ \\
    LES of the transient change of operating condition & $\times$ & $\times$ & - \\
    LES of the steady-state operation& $\times$ & $\times$ & $\times$ \\
   \hline
   \hline
    \end{tabular}
    	\label{tab1}
\end{table*}

Compressible large eddy simulations (LESs) with analytically reduced chemistry (ARC) were performed using the explicit cell-vertex code AVBP \cite{Gicquel2011}. One can refer to \cite{Moureau2005} for the LES equations. The numerical two-step Taylor-Galerkin scheme TTGC \cite{Colin2000} gives third-order accuracy in space and time. The time step is $1\times10^{-7}\,\text{s}$ for atmospheric pressure, and $0.4\times10^{-7}\,\text{s}$ for high pressure (based on the acoustic CFL condition). The Smagorinsky approach \cite{Smagorinsky1963} was used to model the sub-grid Reynolds stress with a constant of 0.18 which has been shown to give good performances in endless studies (e.g. \cite{Colin2000a, Schmitt2007b,SchulzSYMP,SchulzRJICF}). Navier-Stokes characteristic boundary conditions (NSCBC) \cite{Poinsot1992} were imposed at the inlets and the outlet. The the following heat loss formulation was applied to the domain walls: $\dot{q}=(T_{\,\text{wall}}-T_{\,\infty})/R_{\,\text{w}}$, with reference temperature $T_\infty=300\,\text{K}$. The thermal resistances $R_{\,\text{w}}$ are given in Table \ref{tab1}. A priori, we determined wall temperatures at these $R_{\,\text{w}}$ with 2-D, wall-resolved simulations at characteristic velocities, pressures, and temperatures. This resulted in wall temperatures of $700\,\text{K}$ in the mixing zone and $1000\,\text{K}$ in the combustion chamber. We used a coupled velocity/temperature wall-model \cite{VanDriest2003} with the extension for non- isothermal configurations of \citet{Schmitt2007b}. The dynamic thickened flame (DTF) model \cite{Colin2000a} was used for turbulent combustion modeling. The DTF model in combination with ARC has been successfully applied in previous studies with similar operating conditions \cite{Schulz2017,SchulzRJICF,SchulzASME,SchulzSYMP}. \citet{SchulzSYMP} demonstrated that AVBP with ARC and DTF quantitatively captures the different combustion modes (autoignition and propagation) in this sequential combustor, provided that the mesh in regions where  autoignition takes place is sufficiently fine for i) resolving most of the turbulent kinetic energy and ii) for not triggering the DTF. These results were in excellent agreement with experimental hydroxyl planar laser-induced fluorescence (OH-PLIF).

The unstructured computational mesh is refined in the vicinity of the fuel injector, the mixing section, and the flame region. We used Pope's \cite{Pope2000} criterion to determine the quality of our LESs. The criterion reveals that a reliable LES should resolve at least 80\% of the total kinetic energy. Therefore, we computed the ratio between the resolved part of the turbulent kinetic energy and the total kinetic energy a posteriori. Results in \cite{Schulz2018} show that more than 95\% of the total kinetic energy was resolved for the atmospheric case ($\approx 16$ million cells). For the cases at $10\,\text{bar}$, the mesh was significantly refined to resolve at least $80\%$ of the total kinetic energy ($\approx 60$ and $\approx 74$ million cells; see Table \ref{tab1}).

An analytically reduced chemistry (ARC) scheme with 22 transported species was used for simulations at $1\,\text{bar}$. One can refer to \cite{Schulz2017, Schulz2018, Jaravel2018a} for a detailed validation of the scheme. For simulations at $10\,\text{bar}$, a different ARC scheme was obtained from the detailed mechanism GRI-mech 3.0 \cite{gri30}. The directed relation graph with error propagation (DRGEP) method \cite{Pepiot-Desjardins2008} was used to remove 29 species in a first reduction step. Second, the following species were identified as suitable for quasi-steady state (QSS) assumption: HCNN, CH$_2$GSG-CH$_2$, CH$_3$O, HCO, CH$_2$, C$_2$H$_5$, H$_2$O$_2$, and C$_2$H$_3$. Their concentrations are described by analytical expressions \cite{Pepiot-Desjardins2008}. Hence, the following 16 transported species remained in the scheme, named ARC\_10bar: N$_2$, H, H$_2$, O, OH, O$_2$, H$_2$O, HO$_2$, CH$_3$, CH$_2$O, CO$_2$, CO, CH$_3$OH, CH$_4$, C$_2$H$_6$ and C$_2$H$_4$.

\begin{figure}[!t]
	\begin{center}
	\includegraphics[width=66mm]{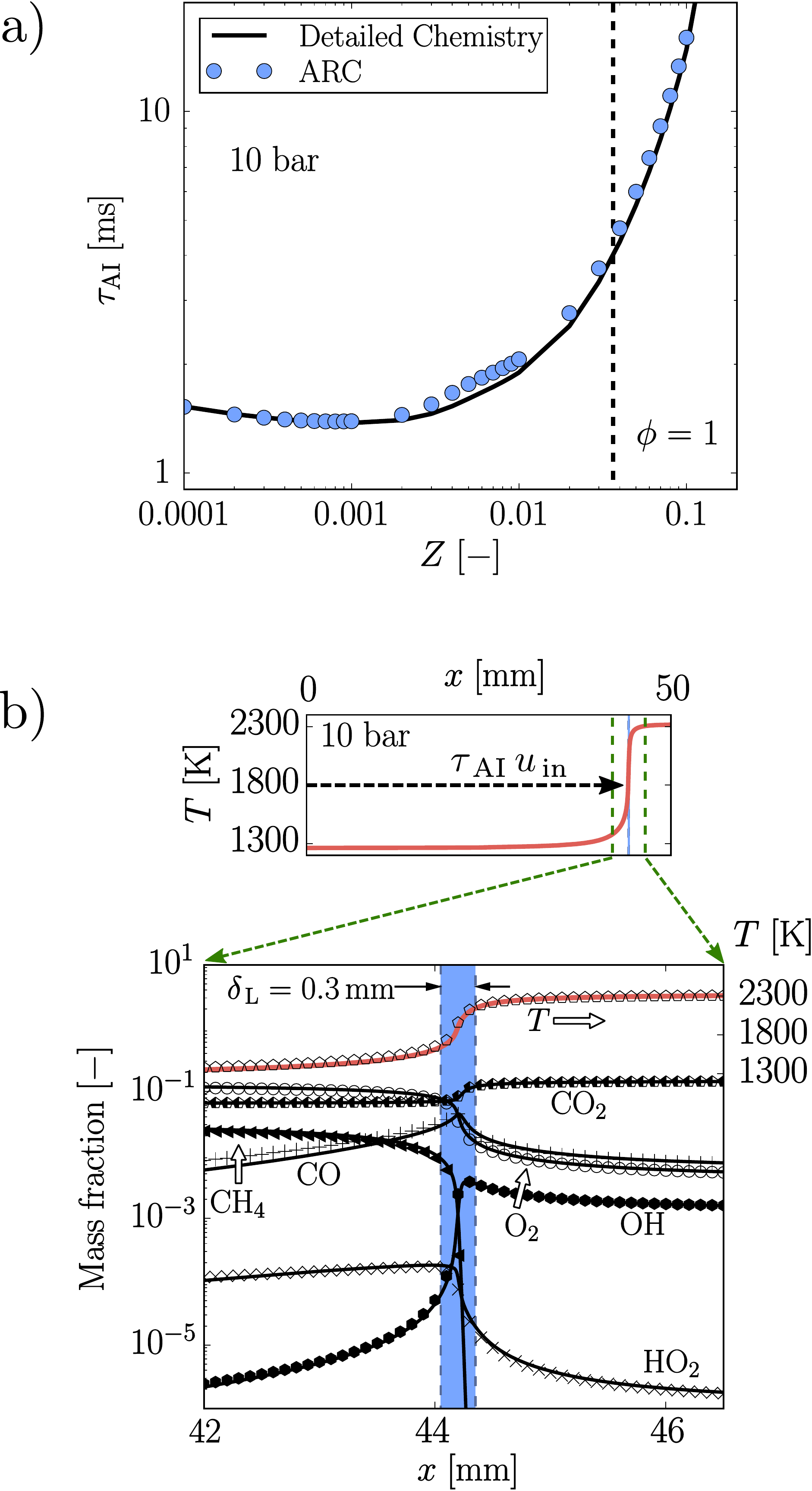}
	\caption{Comparison between analytically reduced chemistry (ARC) mechanism (symbols) and detailed chemistry (lines) at $10\,\text{bar}$. a) Autoignition times $\tau_{\,\text{AI}}$ from 0-D reactor simulations. b): Profiles of temperature and mass fractions of a stoichiometric 1-D autoignition flame. The width of blue region highlights flame thickness $\delta_{\,\text{L}}$. Top part shows the temperature profile with autoignition length $\tau_{\,\text{AI}}\,u_{\,\text{in}}$.}
	\label{fig:chem_stoich}
	\end{center}
\end{figure}
The validation of ARC at $10\,\text{bar}$ in terms of autoignition times (a), and profiles of temperature and species mass fractions for a 1-D autoignition flame (b) is shown in Fig. \ref{fig:chem_stoich}. The lines show results from simulations with Cantera (0-D and 1-D) using GRI-Mech 3.0 \cite{gri30}; results with ARC (symbols) were obtained with 0-D reactor simulations for (a) and a 1-D AVBP simulation for (b). For (b), the inlet velocity $u_{\,\text{in}}$ is set to $10\,\text{m/s}$. An autoignition delay $\tau_{\,\text{AI}}$ of $4.4\,\text{ms}$ results in an autoignition length of $\tau_{\,\text{AI}}\,u_{\,\text{in}}=44\,\text{mm}$. Autoignition times and profiles for temperature and species mass fractions are in excellent agreement with the detailed chemistry.

\subsection{Combustion regimes from 0-D reactors and 1-D flames}
\begin{figure*}
	\begin{center}
	\includegraphics[width=150mm]{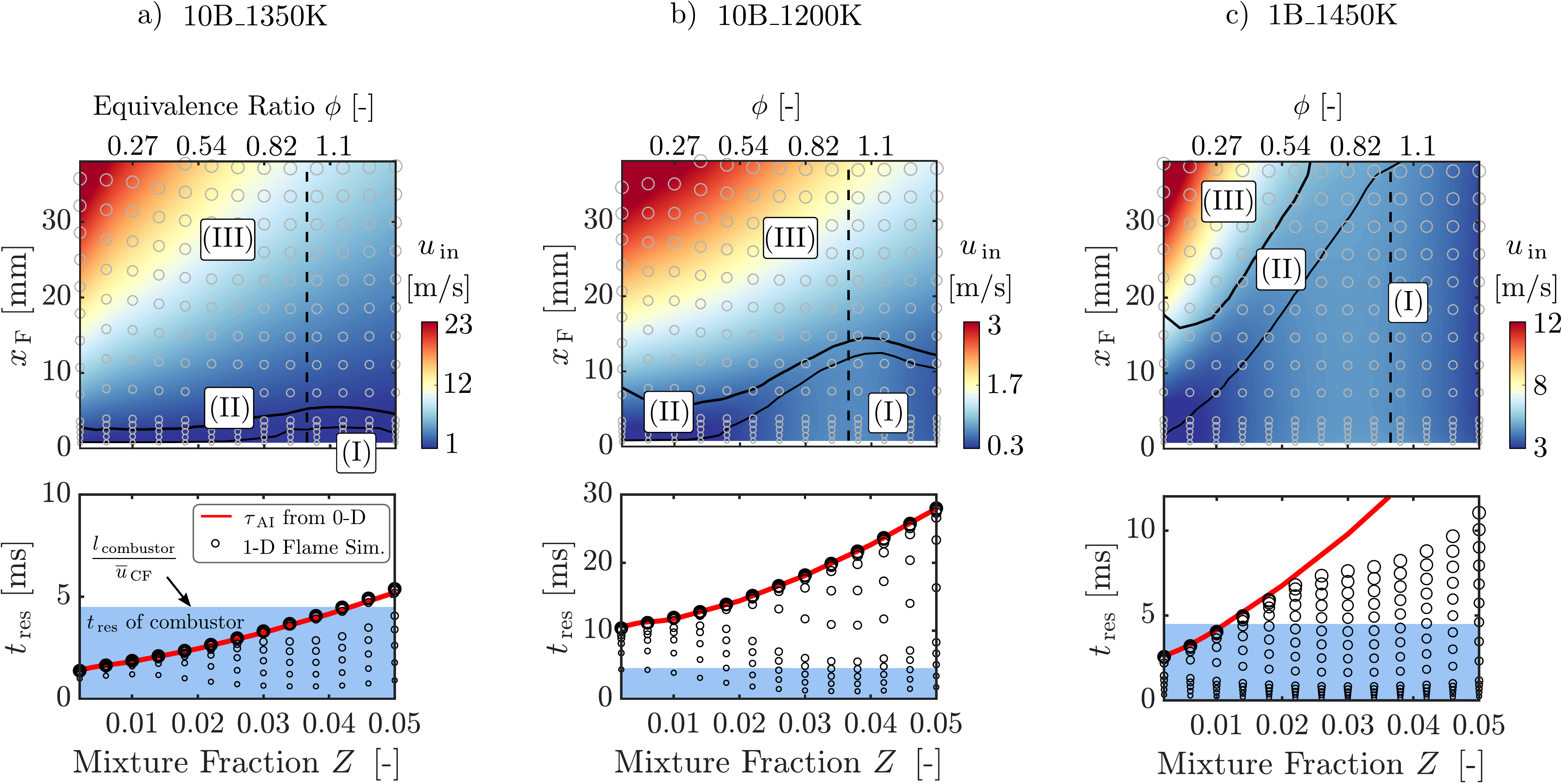}
	\caption{Dominant combustion regimes for the cases (a) to (c). The top row shows combustion regime maps from 1-D simulations similar to Fig. \ref{fig:map1450K}. Bottom plots show the residence time of the reactants $t_{\,\text{res}}$ for 1-D simulations plotted over mixture fraction $Z$. In the top and bottom rows, the diameter of the circles is proportional to the flame position $x_{\,\text{F}}$ allowing the reader to recognize the same circles in the top and bottom. The red line highlights the autoignition delay computed from 0-D reactor simulations. Blue patches show a rough estimation of the residence time range of the combustor. The maximum residence time was computed as $l_{\,\text{combustor}}/\overline{u}_{\text{CF}}$. The combustor length $l_{\,\text{combustor}}$ is defined from the end of the mixing section to the end of the combustion chamber ($280\,\text{mm}$).}
	\label{fig:regimes2}	
	\end{center}
\end{figure*}
Figure \ref{fig:regimes2} investigates the dominant combustion regimes for the three mixtures presented in Table \ref{tab1} from (a) to (c). The top row shows the regime maps as presented in Fig. \ref{fig:regimes1}. From these maps, the  reactants residence times $t_{\,\text{res}}=\int_0^{x_{\,\text{F}}}dx/u$ were computed and shown in the bottom row of Fig. \ref{fig:regimes2}. These maps were obtained from laminar adiabatic perfectly-premixed 1-D flames with detailed chemistry, which is far from the 3-D partially-premixed turbulent configurations investigated in this paper. Nevertheless, it is very informative to compare the residence times $t_{\,\text{res}}$ of the idealized laminar configuration to an estimate of the residence time of the sequential combustor $l_{\,\text{comb}}/\overline{u}_{\,\text{CF}}$, keeping in mind that it is a crude comparison. Many features of the sequential combustor affecting the residence time are not included, for example, recirculation zones, boundary layers, the wake of the jet in crossflow or turbulent fluctuations. Neglecting these possible effects, the blue patch highlights the residence time range of the combustor where the combustor length $l_{\,\text{comb}}$ is defined from the end of the mixing section to the domain outlet ($280\,\text{mm}$).

At $10\,\text{bar}$ and high inlet temperature (10B\_1350K) in (a), mixtures for $Z<0.45$ exhibit autoignition within the residence time of the combustor. Autoignition time-scales were extracted from 0-D reactor simulations and are highlighted by the red solid line. On the contrary, at $10\,\text{bar}$ and decreased inlet temperature (10B\_1250K) in (b), autoignition is not expected to occur. Here, the autoignition times are smaller than the residence time of the combustor for any $Z$. Therefore, flame propagation dominates for (b). At atmospheric pressure (1B\_1450K) in (c), autoignition can occur for very lean $Z<0.1$. For increasing $Z$ the autoignition delay curve is significantly steeper compared to 10B\_1350K in (a). For example, for the global equivalence ratio of case 1B\_1450K ($\phi_{\,\text{g}}=0.76$), it is not expected that autoignition occurs within the residence time of the combustor and that, therefore, flame propagation dominates.

Of course, as already pointed out, these results from 1-D flame solutions and 0-D simulations do not incorporate effects of turbulence. They also do not account for the increased residence times in the wake of jet in crossflow or in the recirculation region of the combustion chamber, which can play an important role during the ignition sequence of the combustor. These points are now scrutinized using data from transient and steady-state turbulent LESs.

\subsection{Transient LESs of ignition sequences} \label{subsec:ign}

\begin{figure}[!t]
	\begin{center}
	\includegraphics[width=88mm]{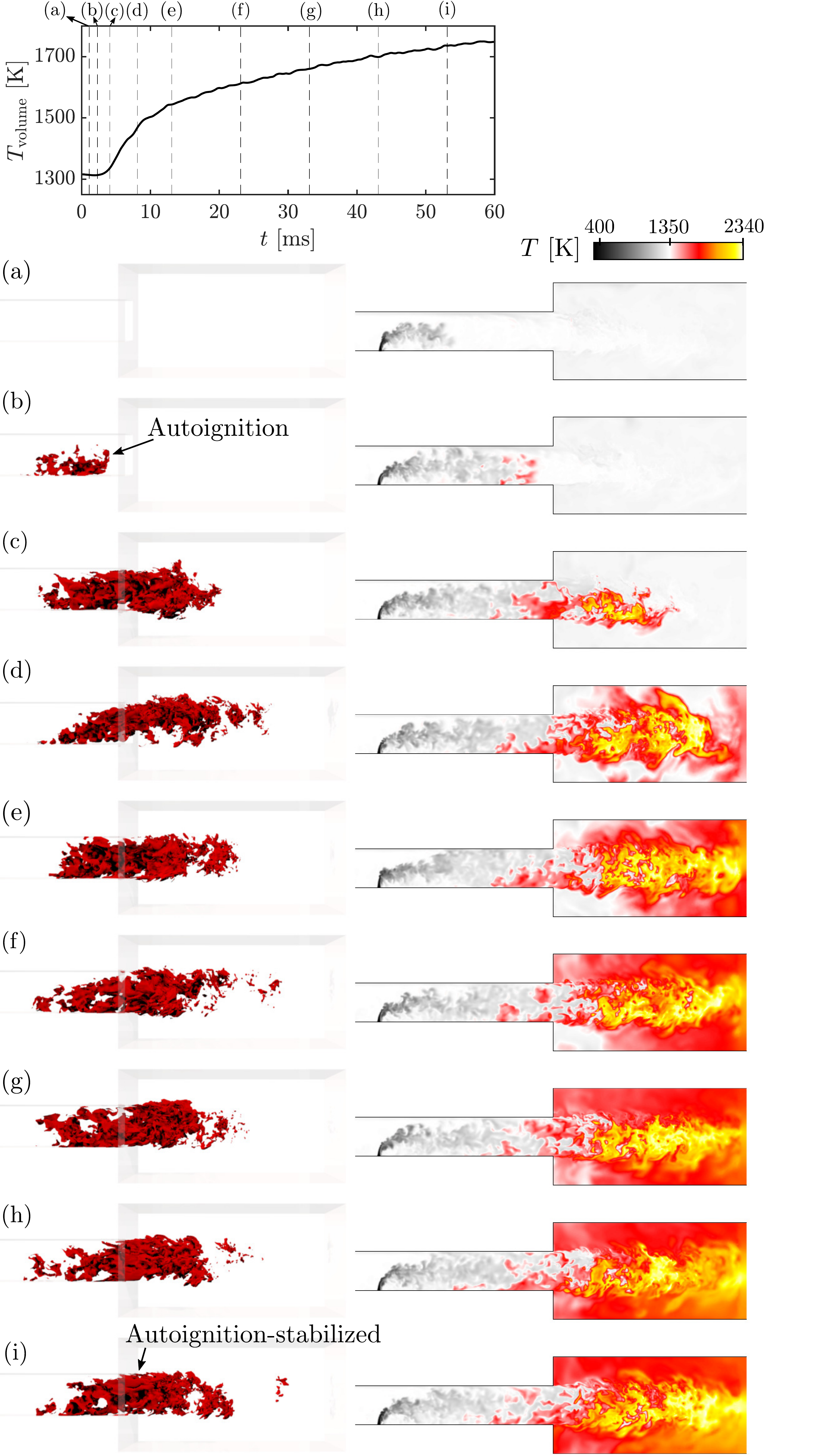}
	\caption{Ignition sequence for case 10B\_1350K with the time trace of volume integrated temperature (top), and instantaneous snapshots of a 3-D rendering of temperature iso-surface at $1500\,\text{K}$ (left column) and temperature contour of a 2-D $y$-$z$-cut through the domain centerline (right column). Fuel injected at $t=0$.  Pressure: $10\,\text{bar}$.}
	\label{fig:ignition_10b}	
	\end{center}
\end{figure}

This subsection investigates transient ignition sequences for these two cases: 10B\_1350K and 1B\_1450K. Figure \ref{fig:ignition_10b} shows the sequence for case 10B\_1350K. Instantaneous snapshots of a 3-D rendering of a temperature iso-surface at $1500\,\text{K}$ (left column) representing the flame front, and the temperature $T$ contour of a 2-D $y$-$z$-cut through the domain centerline (right column) are visualized. The temperature iso-surface is conditioned on fuel mass fraction ($Y_{\,\text{CH}_4}>10^{-4}$) allowing to discard regions with burnt gas temperatures that are reduced due to wall heat losses. Eight subsequent snapshots from (a) to (i) are shown from top to bottom. The instants of time are marked in the top graph, which gives the domain-volume-integrated temperature $T_{\,\text{volume}}$ for each snapshot. The simulation was initialized by the mixture and temperature of the vitiated hos gas inlet. Once the flow without fuel injection reached a fully developed state after approximately two flow-through times ($10\text{ms}$), the simulation was stopped and then restarted with fuel mass flow injection (this instant is defined as $t=0$). Approximately $2\,\text{ms}$ after fuel injection, $T_{\,\text{volume}}$ increases, and autoignition occurs first in the mixing section as shown in (b). At this instant of time, no fuel has reached the combustion chamber. Indeed, the first occurrence of autoignition corresponds roughly to the convective delay of $\approx 2.5\,\text{ms}$ between the fuel injector to the autoignition position, computed with a bulk velocity of $60\,\text{m/s}$. The moment of autoignition onset is in very good agreement with results from 0-D and 1-D simulations (see Fig. \ref{fig:regimes2}a) showing that autoignition occurs first for very lean mixtures after $\approx 1-2\,\text{ms}$. At (c), autoignition occurs also further downstream in the combustion chamber. Here, richer mixtures auto-ignite leading to higher burnt gas temperatures, as seen in the 2-D $T$ contour. From (d) to (i), the overall flame shape does not change anymore: the turbulent flame is anchored in the mixing section and extends axially into the combustion chamber, as seen in the 3-D snapshots. The 2-D cuts show that the flame also extends over the whole mixing section height ($y$-direction). The same was observed for the $z$-direction (not shown). Autoignition is initiated first in the bottom of the mixing section due to the presence of regions with lower velocities in the wake of the jet and therefore longer residence times. From (d) to (i), merely the recirculation zone fills with hot burnt gases, as visualized in the 2-D plots.

\begin{figure}[!t]
	\begin{center}
	\includegraphics[width=88mm]{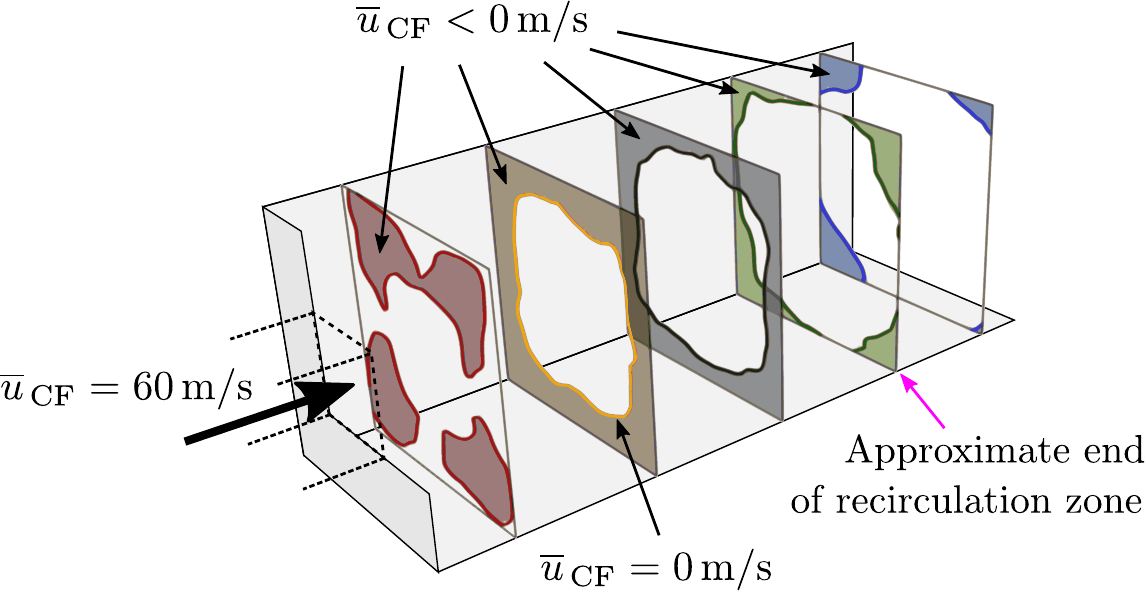}
	\caption{Time-averaged shape of the recirculation zone in the combustion chamber obtained from a steady-state simulation of case 1B\_1450K. The $x$-$y$-cuts are located at the following distances with respect to the combustion chamber inlet: 30, 90, 150, 210 and $270\,\text{mm}$. The approximate length of the recirculation zone is defined and highlighted at $210\,\text{mm}$.}
	\label{fig:recirc}	
	\end{center}
\end{figure}
The recirculation zones play an important role for flame stabilization in the ignition sequence at atmospheric pressure (case 1B\_1450K). The time-averaged shape is, therefore, obtained from a steady-state simulation of case 1B\_1450K, and visualized in several $x$-$y$-cuts at 30, 90, 150, 210 and $270\,\text{mm}$ with respect to the combustion chamber inlet in Fig. \ref{fig:recirc}. The darker areas in the $x$-$y$-cuts highlight regions with negative axial velocity $\overline{u}_{\,\text{CF}}$. Close to the combustion chamber inlet, the regions with negative velocity are disconnected which can be attributed to the 3-D flow effects and the interaction between the four recirculation zones. However, the present objective is not to provide an analysis of the recirculation zones flow dynamics but rather to give an estimate of the length of the recirculation zones in axial direction. Although, there are still small regions with negative $\overline{u}_{\,\text{CF}}$ in the corners of the combustion chamber in the last cut at $270\,\text{mm}$, the approximate length of the recirculation zone is defined at $210\,\text{mm}$ with respect to the combustion chamber inlet and is highlighted in Fig. \ref{fig:recirc}.

\begin{figure}[!t]
	\begin{center}
	\includegraphics[width=88mm]{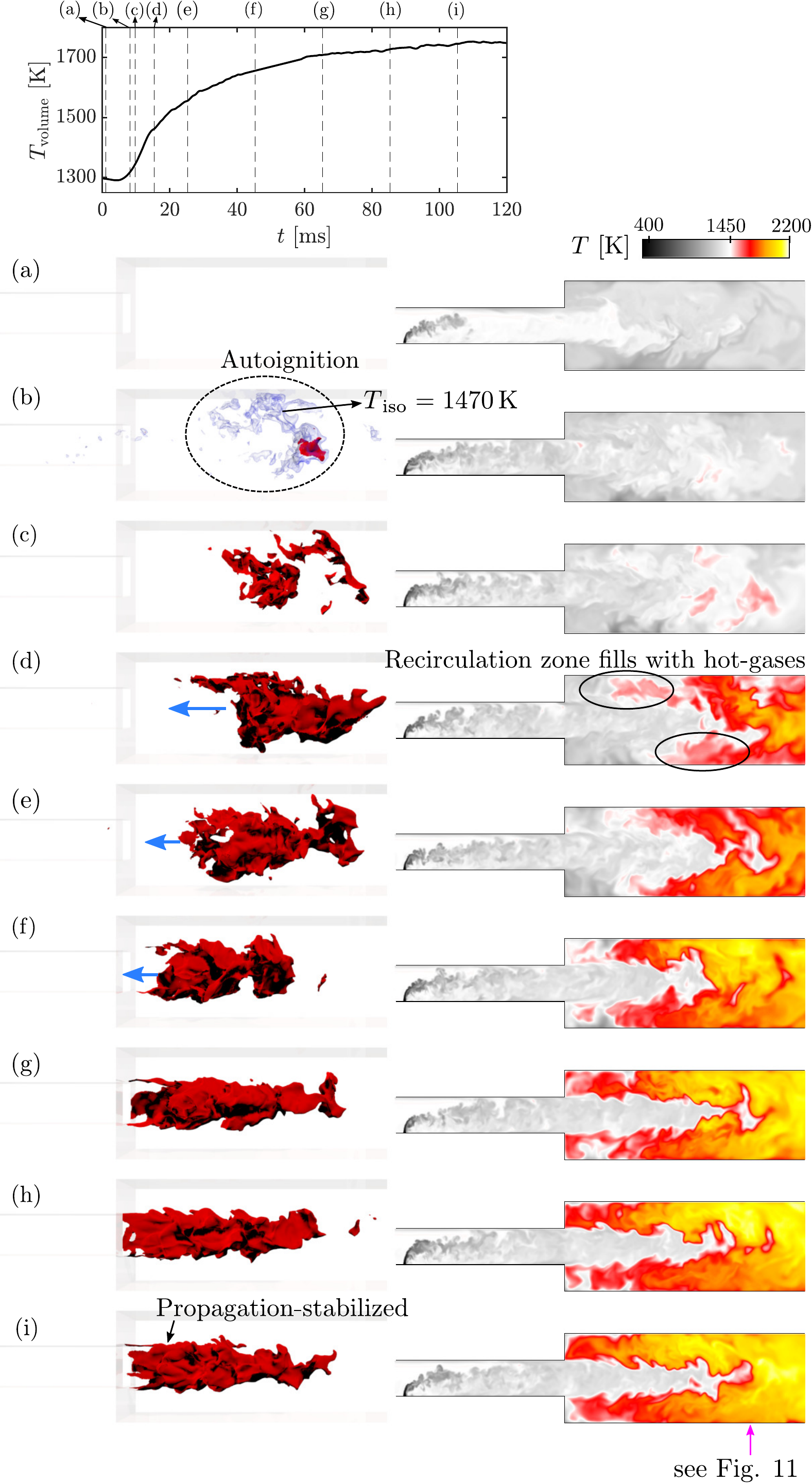}
	\caption{Ignition sequence for case 1B\_1450K with the time trace of volume integrated temperature (top), and instantaneous snapshots of a 3-D rendering of temperature iso-surface at $1500\,\text{K}$ in red (left column) and temperature contour of a 2-D $y$-$z$-cut through the domain centerline (right column). In (b), the temperature iso-surface at $1470\,\text{K}$ is also shown. Fuel injected at $t=0$. Pressure: $1\,\text{bar}$.}
	\label{fig:ignition_1b}	
	\end{center}
\end{figure}
\begin{figure*}[!ht]
	\begin{center}
	\includegraphics[width=120mm]{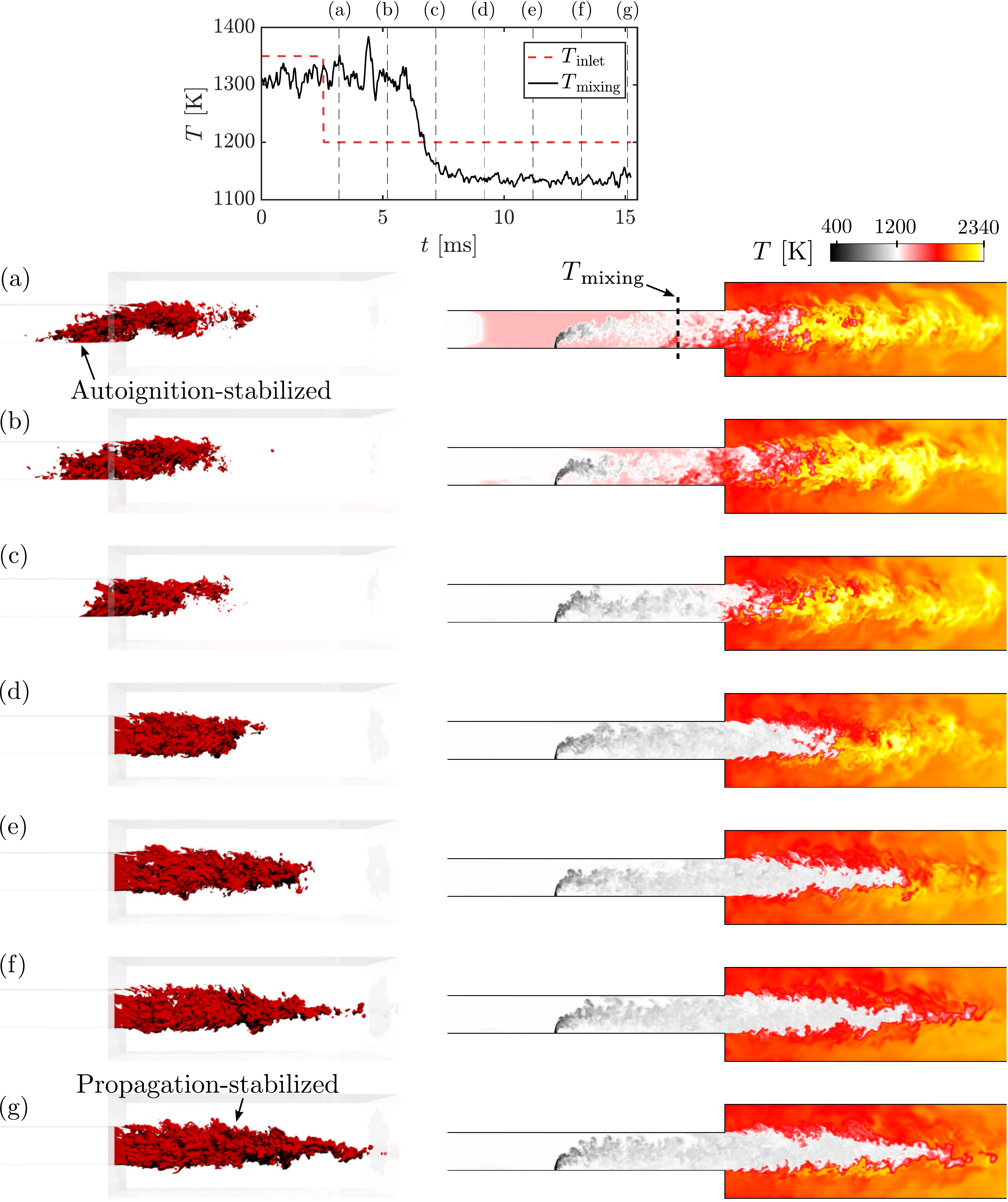}
	\caption{Transient change of operating condition from 10B\_1350K to 10B\_1200K (vitiated hot gas temperature decreased by $150\,\text{K}$). Left and right columns show 3-D rendering of temperature iso-surface at $1500\,\text{K}$, and temperature contour of a 2-D $y$-$z$-cut through the domain centerline. Top time trace shows the temperature at the domain inlet $T_{\,\text{in}}$ and in the mixing section $T_{\,\text{mixing}}$ (spatial average on an $x$-$y$-cut marked in (a)). Pressure: $10\,\text{bar}$.}
	\label{fig:transient}	
	\end{center}
\end{figure*}

The ignition sequence at atmospheric pressure is shown in Fig. \ref{fig:ignition_1b}. The left column visualizes the flame front with 3-D rendering of a temperature iso-surface at $1500\,\text{K}$ (conditioned on $Y_{\,\text{CH}_4}>10^{-4}$). The right column shows the $T$ contour of the 2-D centerline $y$-$z$-cut and, in snapshot (i), the approximate spread of the recirculation zone presented in Fig. \ref{fig:recirc}. The snapshots are marked in the top volume-integrated temperature $T_{\,\text{volume}}$ time trace. The simulation was initialized as in the high-pressure case 10B\_1350K (see Fig. \ref{fig:transient}) (but with a temperature of $1450\,\text{K}$) and again after two flow-through times, fuel was injected; this instant is again defined as $t=0$. After $\approx 7-8\,\text{ms}$, a $1500\,\text{K}$ iso-surface is observed in the combustion chamber for the first time, as seen in (b). An iso-surface at $1470\,\text{K}$ is also shown (light blue), and corresponds to autoignition of very lean mixture with low heat release rate and therefore a low increase of the temperature (20 K for that isosurface). The drop of the volume-integrated temperature $T_{\,\text{volume}}$  induced by the injection of cold fuel at the beginning of the sequence (during the first 4-5 ms) is followed by a steady rise of the temperature due to the autoignition process in the chamber. The regions with moderate $T$ are mainly located in the recirculation zone of the combustion chamber. This is in agreement with results from 0-D and 1-D simulations (see Fig. \ref{fig:regimes2}c) showing that the smallest autoignition times $\tau_{\,\text{AI}}$ are found for very lean conditions (very small $Z$), and that the $\tau_{\,\text{AI}}$ curve is significantly steeper for increasing $Z$ compared to, for example, Fig. \ref{fig:regimes2}a. Therefore, the residence time of the ignitable mixture needs to be sufficiently large to initiate autoignition with increased heat release rates, which is achieved in the recirculation zone, as shown in the 3-D LES. Without recirculation zone, we would expect a ``no ignition" regime \cite{Markides2005}. For (c), the flame front spreads and one observes a transition to a ``lifted" flame (see (d)) that propagates against the incoming flow and whose stabilization is supported by the hot gases in the downstream part of the recirculation zone. From (d) to (f), also the upstream part of the recirculation zone gradually fills with hot gases (see 2-D contours) allowing the flame stabilization position to move towards the combustion chamber inlet, as highlighted with the blue arrows in the 3-D plots. Finally, from (g) to (i), the recirculation zone is almost completely filled with burnt hot gases leading to a flame that is anchored very close to the combustion chamber inlet. Here, the stabilized turbulent propagating flame has a typical conical shape.

These transient simulations give substantial insight into the dominant combustion regimes that are responsible for ignition and subsequent flame anchoring. For the high-pressure case (10B\_1350K), the fuel hot gas mixture auto-ignites in the mixing section and the flame stabilizes due to continuous autoignition of the mixture. The reaction zone extends over the whole mixing section height and width into the combustion chamber. The most upstream positions of ``first" autoignition and subsequent flame stabilization are almost identical and do not change on average. For the atmospheric case (1B\_1450K), the mixture auto-ignites in the recirculation zone, followed by a transition to a propagating ``lifted" flame. Subsequently, the flame stabilizes close to the combustion chamber inlet in presence of hot burnt gases in the recirculation zone. These recirculating burnt gases determine the transient position of the propagating flame during the stabilization process. 

\subsection{Transient LES of a changing operating condition} \label{sec:transient}
Figure \ref{fig:transient} presents instantaneous snapshots from (a) to (g) of a transient 3-D simulation with changing operating conditions from case 10B\_1350K to 10B\_1200K. The simulation was started from a solution at steady-state operation of case 10B\_1350K; the instant of start is defined as $t=0$. After $2.5\,\text{ms}$, the simulation was stopped and the temperature of the vitiated hot gas inlet was decreased by $150\,\text{K}$; then the simulation was restarted, as seen in the top time trace of $T_{\,\text{in}}$. This plot also shows the temperature in the mixing section extracted $50\,\text{mm}$ upstream of the combustion chamber inlet,  marked in (a). The snapshots (marked in the time plot) in the left and right column show the 3-D rendering of a temperature iso-surface at $1500\,\text{K}$ (conditioned on $Y_{\,\text{CH}_4}>10^{-4}$), and the $T$ contour of the 2-D centerline $y$-$z$-cut, respectively. The steady-state flame of case 10B\_1350K in the snapshot (a) is stabilized by continuous autoignition of the mixture of the jet and the vitiated hot gas, as shown in Fig. \ref{fig:ignition_10b}. When the vitiated gases with decreased $T$ reach the jet in crossflow mixing section (b), the mixture does not auto-ignite anymore (c) and the reaction zone gets pushed into the combustion chamber, as seen from (c) to (d). Subsequently, a turbulent flame with a classical cone shape develops, shown from (d) to (e). Snapshots (f) and (g) show a typical propagating type flame front which is significantly more wrinkled due to the increased turbulent Reynolds number ($Re_{\,\text{t}}\approx 1006$), and hence, higher flame surface density \cite{Lachaux2005} compared to the atmospheric case in Fig. \ref{fig:ignition_1b} ($Re_{\,\text{t}}\approx 38$). It is also shown that the flame angle decreases leading to an increase of the flame length compared to 1B\_1450K. These findings are in very good agreement with results from 0-D and 1-D simulations (see Fig. \ref{fig:regimes2}b), identifying flame propagation as the dominant combustion regime for 10B\_1200K.

\subsection{Comparison of steady-state LESs}
\begin{figure*}[!t]
	\begin{center}
	\includegraphics[width=160mm]{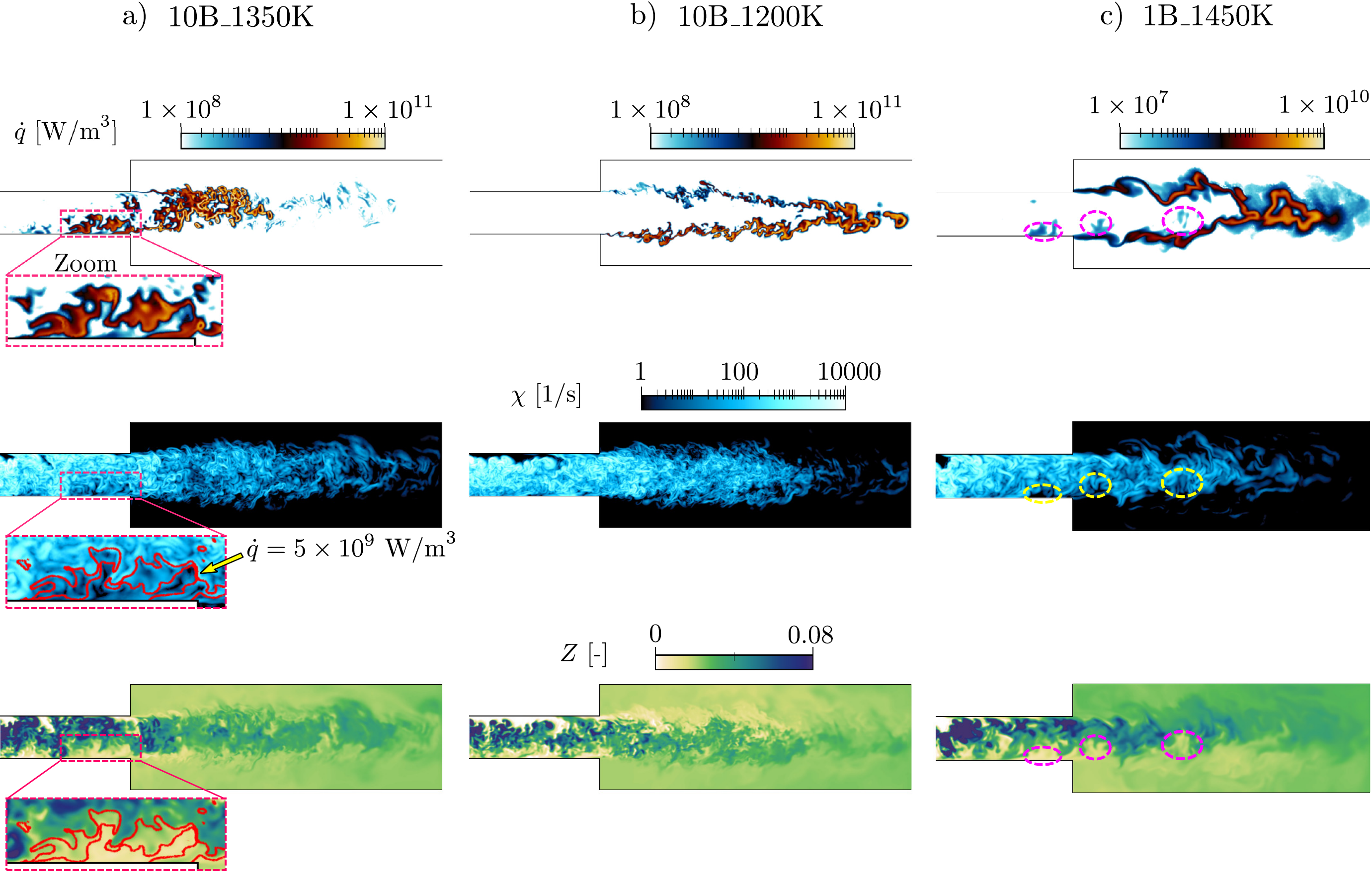}
	\caption{Typical instantaneous snapshots of the 2-D centerline $y$-$z$-cut for the three operating conditions at steady-state operation from (a) to (c). These contours are shown from top to bottom: heat release rate $\dot{q}$, scalar dissipation rate $\chi$ and mixture fraction $Z$. Zoom in (a) shows conditions at the upstream autoignition position in the mixing section ($\chi$ and $Z$ fields overlaid with $\dot{q}$ iso-line). For the propagating flame in (b), no upstream autoignition events are observed. Ellipses in (c) show favorable conditions for autoignition (low $\chi$ and low $Z$) upstream of the stabilized flame. Low heat release rate autoignition events occur occasionally at such favorable conditions.}
	\label{fig:compare_inst}	
	\end{center}
\end{figure*}
Figure \ref{fig:compare_inst} compares typical instantaneous snapshots of the 2-D centerline $y$-$z$-cut for the three operating conditions at steady-state operation. Three properties are visualized from top to bottom: heat release rate $\dot{q}$, scalar dissipation rate $\chi$ (both in logarithmic scale), and mixture fraction $Z$. The definition $2D\left|\nabla Z\right|^2$ with mass diffusivity $D$ was chosen for $\chi$. The zoom with an overlaid $\dot{q}$ iso-line at $5\times10^9 \,\text{W/m}^3$ visualizes the conditions in the most upstream region of the autoignition stabilized flame in (a). It is shown that regions with heat release rates coincide with relatively low $\chi$ and $Z$, which is characteristic for autoignition in turbulent mixing flows \cite{Mastorakos1997}. For (b), regions with relatively low $\chi$ and $Z$ exist, as seen for example in the bottom part of the mixing section. Nevertheless, no autoignition events are observed upstream of the stabilized flame, as seen in the $\dot{q}$ contour. For (c), three regions with low $\chi$ and $Z$ are highlighted with ellipses. In these regions autoignition occurs at heat release rates that are approximately two orders of magnitudes lower than the ones observed in the stabilized propagating flame. This is in very good agreement with results presented in \cite{Schulz2018}. In \cite{Schulz2018}, the authors observed coherent occurrences of autoignition kernels upstream of the propagating flame for harmonic oscillations of the vitiated crossflow temperature. 
\begin{figure}[!t]
	\begin{center}
	\includegraphics[width=86mm]{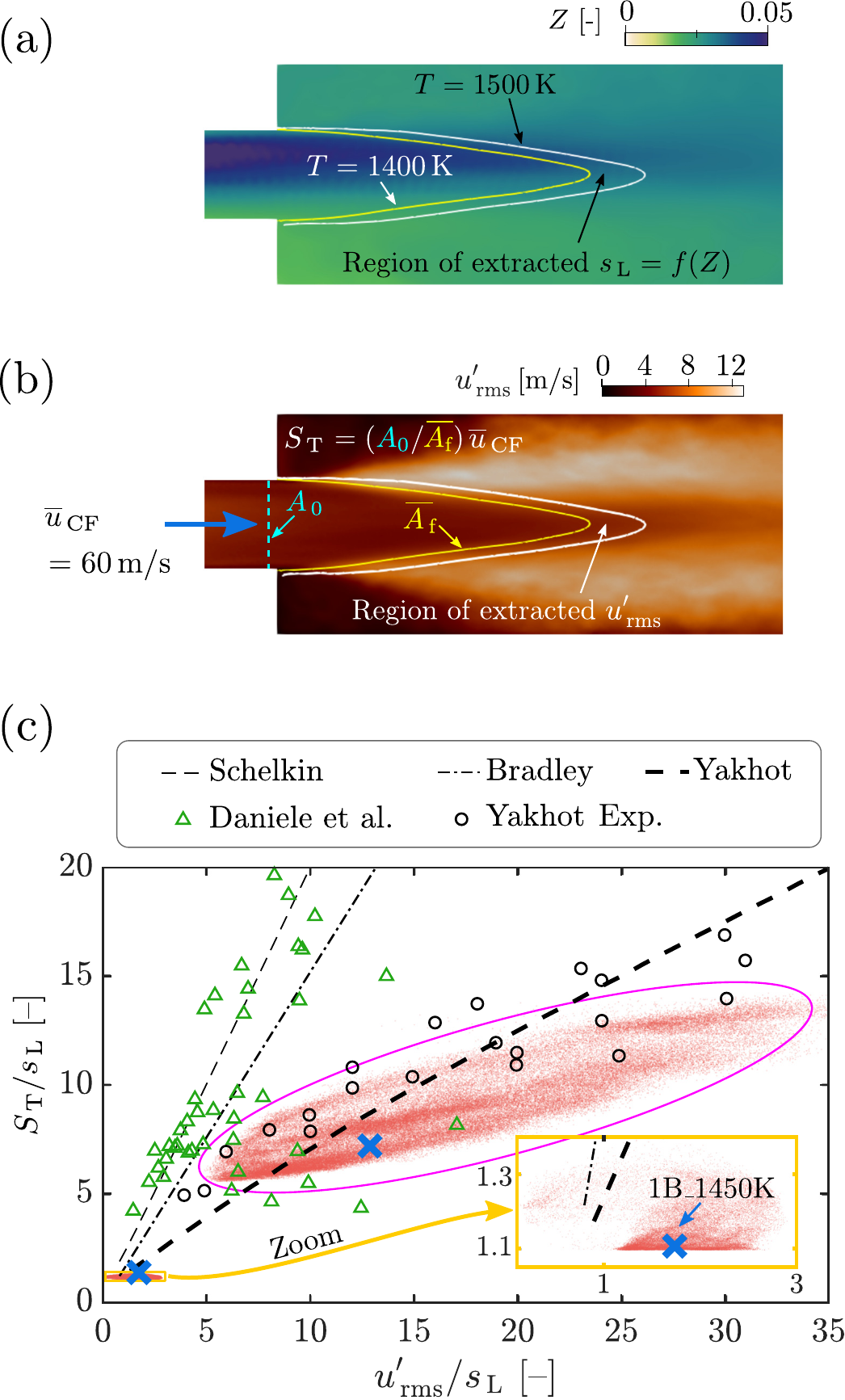}
	\caption{(a): 2-D $y$-$z$-cut of time-averaged mixture fraction $Z$ contour used to extract the local $s_{\,\text{L}}$. (b): 2-D $y$-$z$-cut of turbulent intensity $u'_{\,\text{rms}}$ contour. $s_{\,\text{L}}$ and $u'_{\,\text{rms}}$ were extracted in each point of a 3-D flame region conditioned on $1400\leq T\leq 1500$ (here, shown in 2-D for case 1B\_1450K). (c): Scatter plot of normalized global turbulent flame speed $S_{\,\text{T}}$ over normalized $u'_{\,\text{rms}}$ for cases 1B\_1450K (zoom) and 10B\_1200K (ellipse). Both cases are partially-premixed. Lines give correlations of \citet{Shelkin1943}, \citet{Bradley1992} and \citet{Yakhot1988}; symbols show experimental data of \citet{Daniele2013} and \citet{Yakhot1988} (all perfectly-premixed). Crosses at $S_{\,\text{L}}$ of global $Z$, and turbulent intensity averaged in the flame region ${<}u'_{\,\text{rms}}{>}$. ${<}u'_{\,\text{rms}}{>}$ for 1B\_1450K and 10B\_1200K are $7.8$ and $8\,\text{m/s}$.}
	\label{fig:turb_vel}	
	\end{center}
\end{figure}
A large number of autoignition kernels that were characterized by elevated heat release rates were observed in the advected \emph{hot} streamwise strata. They led to flame front merging and, therefore, strong heat release rate fluctuations in the combustion chamber. Simulations of this work feature a constant vitiated hot gas temperature and local autoignition events were only observed occasionally. They occurred preferably downstream of the wake of the jet in crossflow as seen in (c) and they did not induce a merging of the propagating flame front during the entire simulation. This is also in very good agreement with 0-D and 1-D results (see Fig. \ref{fig:regimes2}c) showing that autoignition times can be smaller than the maximum combustor residence time for very lean mixtures (small $Z$). These predictions also show that autoignition delays are significantly bigger than the combustor residence time for richer mixtures with relatively high heat release rates due to a very steep $\tau_{\,\text{AI}}$ curve.

Figure \ref{fig:turb_vel} compares the normalized global turbulent flame speed of this work (partially-premixed) with correlations and experimental data from the literature (perfectly-premixed). The global turbulent flame speed was computed as $S_{\,\text{T}}=(A_{\text{0}}/\overline{A_{\text{f}}})\,\overline{u}_{\,\text{CF}}$, as done, for example, in \cite{Daniele2013}. This formulation was obtained from a continuity analysis stating that the mass flow is conserved through the mixing section surface $A_{\text{0}}$ ($40\times40\,\text{mm}$) and the flame front surface $\overline{A_{\text{f}}}$. Here, $\overline{A_{\text{f}}}$ is the time-averaged area of a temperature iso-surface at $1400\,\text{K}$. This temperature corresponds to a progress variable of $0.1$. Figure \ref{fig:turb_vel}a shows the 2-D centerline $y$-$z$-cut of the time-averaged mixture fraction $Z$ for case 1B\_1450K. The mixture fraction and, therefore, the local laminar flame speed $s_{\,\text{L}}$ are not constant over the flame front; they range from $0.013\leq Z\leq 0.038$ and $3.6\leq s_{\,\text{L}} \leq 4.5\,\text{m/s}$ for 1B\_1450K; and $0.01\leq Z\leq 0.032$ and $0.32\leq s_{\,\text{L}} \leq 0.8\,\text{m/s}$ for 10B\_1350K. Figure \ref{fig:turb_vel}b shows the contour of the turbulent intensity $u'_{\,\text{rms}}$ in the same cut for case 1B\_1450K. For the configurations in this work, $u'_{\,\text{rms}}$ is also not constant over the flame front, which is in contrast to \cite{Daniele2013}. The range of $u'_{\,\text{rms}}$ and the spatially-averaged value ${<}u'_{\,\text{rms}}{>}$ are comparable for both cases (1B\_1450K and 10B\_1200K) in the flame front. The reader should be aware that $u'_{\,\text{rms}}$ is the turbulent intensity that was resolved in the LESs and, therefore, it depends on the grid size. Although the mesh has a very good resolution in the region used for data extraction, there is still a contribution of the unresolved $u'_{\,\text{rms}}$ which is not accounted for in Fig. \ref{fig:turb_vel}. Local $s_{\,\text{L}}$ and $u'_{\,\text{rms}}$ were extracted in the 3-D flame front region shown in Fig. \ref{fig:turb_vel}a and b.

In Fig. \ref{fig:turb_vel}c, the global turbulent flame speed is normalized by the local $s_{\,\text{L}}$ in each point, and is plotted against the local $u'_{\,\text{rms}}$ which is also normalized by the local $s_{\,\text{L}}$. The scatter plot shows a wider distribution of points for case 10B\_1350K compared to 1B\_1450K. The range of $u'_{\,\text{rms}}$ is comparable for both cases, and therefore, this difference is attributed to the wider range of the laminar flame speed $s_{\,\text{L}}$ for 10B\_1350K. A finer or coarser mesh resolution changes the amount of the resolved turbulent kinetic energy and, therefore, it is expected that this can induce a shift of the scatter along the $x$-axis. Nevertheless, it is not expected that this will have a significant effect on the difference between the scatter distribution of both cases which is driven by the inhomogeneous mixture fraction and, therefore, the laminar flame speed distribution in the flame region. Both cases agree best with the correlation and the experiments presented by \citet{Yakhot1988}. The deviation of points from \cite{Yakhot1988} for higher $u'_{\,\text{rms}}/s_{\,\text{L}}$, and the deviation from correlations of \citet{Shelkin1943} and \citet{Bradley1992} could be explained by the strong mixture fraction fluctuations, as seen in the instantaneous $Z$ snapshots in Fig. \ref{fig:compare_inst}. The average root-mean-square mixture fraction fluctuation at the combustion chamber inlet is ${<}Z_{\,\text{rms}}{>}\approx 0.01$, corresponding to $\approx40\%$ of the mean $Z$ value. \citet{Garrido-Lopez2005} showed that for $u'_{\,\text{rms}}/s_{\,\text{L}}>1$, increasing levels of $Z$ inhomogeneity lead to a reduction of the turbulent flame speed. They considered $Z_{\,\text{rms}}$ up to $30\%$ of the mean $Z$ value leading to a decrease of the burning rate up to 25\%.

\begin{figure*}[!t]
	\begin{center}
	\includegraphics[width=160mm]{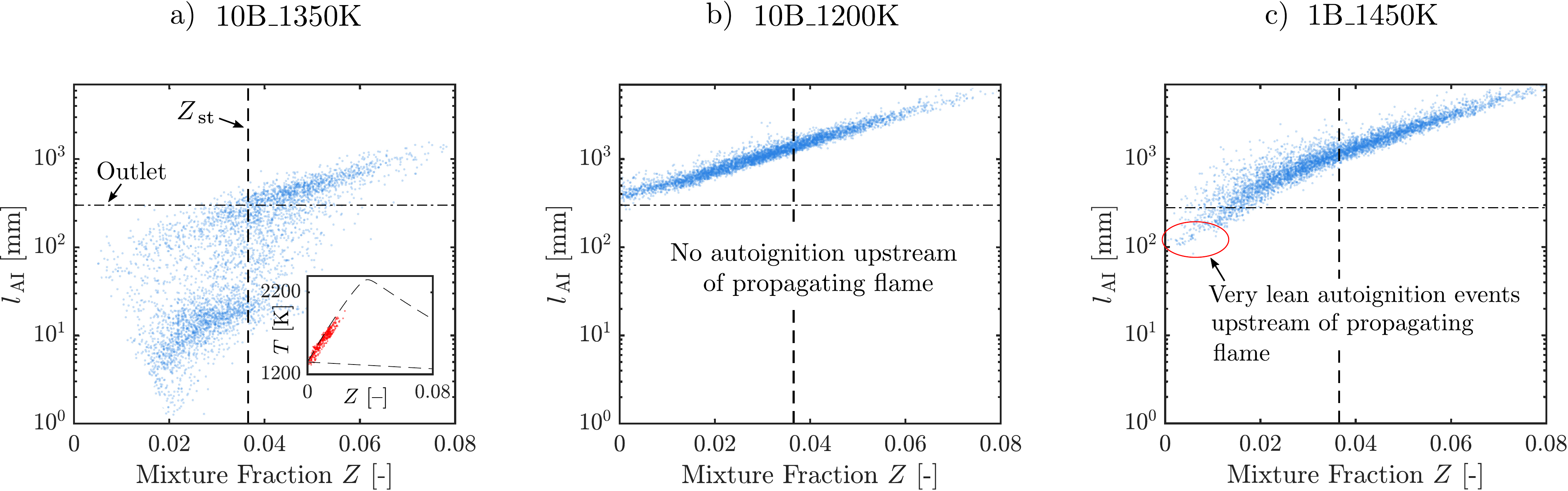}
	\caption{Scatter plot of autoignition length $l_{\,\text{AI}}$ (logarithmic scale) over mixture fraction $Z$ computed from 3-D LESs data extracted at the combustion chamber inlet  cross section ($x-y$ plane). The dashed horizontal line marks the position of the domain outlet. The vertical dashed line at stoichiometric $Z$. $l_{\,\text{AI}}$ computed with the product of axial velocity $u$ and autoignition time $\tau_{\,\text{AI}}$ at each point. $\tau_{\,\text{AI}}$ computed with 0-D reactor simulations. Approximately $4500$ points were extracted for each case. Inset in (a) shows points in thermo-chemical equilibrium.}
	\label{fig:scatter_PSR}	
	\end{center}
\end{figure*}
Figure \ref{fig:scatter_PSR} shows scatter plots of the autoignition length $l_{\,\text{AI}}$ (logarithmic scale) over mixture fraction $Z$ obtained from 3-D LESs for the three operating conditions (a) to (c). Data were extracted from equidistant points distributed at the combustion chamber inlet  cross section ($x-y$ plane). At each point, the autoignition length was computed with $l_{\,\text{AI}}=u\,\tau_{\,\text{AI}}$ using the local axial velocity $u$ and the autoignition time $\tau_{\,\text{AI}}$. The autoignition time was computed with the local temperature $T$ and local mixture composition obtained from the mixture fraction $Z$. The minimum autoignition length $l_{\,\text{AI}}=0$ is defined at the axial position used for data extraction (the end of the mixing section); the horizontal dashed line marks the distance between the data extraction plane and the domain outlet. The vertical dashed line highlights the stoichiometric $Z$. The aim of this figure is to compare the order of magnitude of expected autoignition positions and compare them with results from 0-D and 1-D Cantera simulations. The effect of scalar dissipation rates fluctuations, which can be important for autoignition in turbulent conditions \cite{Mastorakos1997}, is neglected.

For 10B\_1350K in (a), $l_{\,\text{AI}}$ are significantly shorter than the domain outlet for $Z<0.35$. In fact, many points already reached thermal equilibrium; they were discarded for the calculation of $l_{\,\text{AI}}$, and they are highlighted in the inset plot of (a). The inset shows that these points already transitioned from “pure mixing” (bottom dashed line) to complete reaction at thermo-chemical equilibrium (top dashed line). These burnt gases can heat up their neighboring reactants due to turbulent mixing and diffusion of energy and radicals \cite{SchulzRJICF}. This is also promoted by the relatively flat $\tau_{\,\text{AI}}$ curve, shown in Fig. \ref{fig:regimes2}a. Hence, it is expected that autoignition lengths of higher mixture fractions shift towards smaller $l_{\,\text{AI}}$ and smaller $Z$ while being advected downstream. On the contrary, for (b), no autoignition events are expected to occur before the combustor outlet. For (c), autoignition events can occur at very small mixture fractions. However, most of the points are located at axial locations that are larger than the outlet position. The relatively steep $\tau_{\,\text{AI}}$ curve in Fig. \ref{fig:regimes2}c indicates that very lean mixtures can indeed auto-ignite. However, for larger $Z$, autoignition delays and therefore, $l_{\,\text{AI}}$ increase significantly.

\section{Conclusion}
The paper investigates the dominant combustion regimes in the second stage of a partially-premixed sequential combustor with a jet in crossflow fuel injection into vitiated hot gas and flame stabilization sufficiently downstream of the fuel injection.\\
The first part of this work investigates the transition from autoignition to flame propagation at elevated temperatures and atmospheric pressure with 1-D flame simulations. These simulations were performed for mixtures of methane, and vitiated gas at $1000$ to $1600\,\text{K}$. The examined conditions are relevant for the operation of the second stage of sequential combustors. Steady-state 1-D simulations with varying flame positions and mixture compositions identify the boundaries between three combustion regimes: (i) autoignition, (ii) flame propagation assisted by autoignition, and (iii) pure flame propagation. The effect of varying the temperature of the reactants on the three regimes is also investigated. Moreover, a transient 1-D flame simulation with an inlet mixture at $1400\,\text{K}$ was performed. Results show that the mixture auto-ignites and subsequently, a flame propagates towards the inlet of the computational domain. All of the three combustion regimes are present in this simulation.\\
This work also discusses the applicability of the chemical explosive mode analysis (CEMA) to distinguish autoignition and flame propagation. Results show that a chemical explosive mode (CEM) is detected for the propagating flame with hot reactants and the autoignition flame, both at characteristic conditions of the present configuration. Indeed, both flames eventually auto-ignite. However, a propagating flame with CEM in the unburnt reactants can evolve after autoignition.

Then, large eddy simulations (LESs) with analytically reduced chemistry (ARC) were performed for three operating conditions. The combustion regime maps were used to understand the combustion modes at play in these turbulent LESs. First, a comparison of the ignition sequences for two operating conditions is presented: one with autoignition as the dominant combustion regime ($10\,\text{bar}$) and one that is initiated by autoignition but that transfers to a flame stabilized by propagation ($1\,\text{bar}$). For the latter one, the mixture auto-ignites in the recirculation zone where residence time of the ignitable mixture are sufficiently large to initiate autoignition with increased heat release rates. Subsequently, the flame transitions to a ``lifted" propagating flame that moves towards to combustion chamber inlet. The recirculating burnt gases determine the transient position of the flame. Finally, in the steady-state, it anchors close to the combustion chamber inlet. On the contrary, for the autoignition dominated flame, the recirculation zone does not play an important role for flame stabilization. Here, the flame stabilizes due to continuous autoignition of the mixture with a reaction zone that extends over the whole mixing section height and width into the combustion chamber. The position of ``first" autoignition and subsequent flame stabilization does not change on average. 0-D and 1-D predictions show a relatively flat autoignition curve which benefits the continuous self-ignition over the relevant mixture fraction range.\\
Second, the inlet temperature of the autoignition flame ($10\,\text{bar}$) is decreased by $150\,\text{K}$ in another transient LES. This decrease results in a change of the dominant combustion regime from autoignition to flame propagation. This is in very good agreement with the results from 0-D and 1-D flame simulations. For both propagating flames, the global turbulent flame speeds were computed and compared to experiments and correlations (both perfectly-premixed) from the literature. We obtained local laminar speeds and local turbulent intensities over the flame fronts and plotted these points into a turbulent flame speed diagram. The laminar flame speed is not constant over the flame front due to partially-premixed conditions. This results in a wide distribution of points which is particularly large for $10\,\text{bar}$ due to a wider range of the laminar flame speed compared to the atmospheric pressure flame. 

\section*{Acknowledgments}\noindent
This study is supported by the Swiss National Science Foundation under grant 160579 and the Swiss National Supercomputing Centre under grant s685. We gratefully acknowledge CERFACS for providing AVBP; especially thanks to G. Staffelbach for the technical support. We also thank P. Pepiot for providing YARC. 

\bibliographystyle{model1-num-names}

\end{document}